\documentclass[12pt,preprint]{elsarticle}
\usepackage{amsmath}
\usepackage{newtxmath}
\usepackage[T1]{fontenc}
\usepackage[latin9]{inputenc}
\usepackage{graphicx}
\usepackage[bookmarks=false,
 breaklinks=false,pdfborder={0 0 1},backref=section,colorlinks=false]
 {hyperref}
\hypersetup{pdftitle={Dislocation correlations in GaN epitaxial films revealed by EBSD and XRD},
 pdfauthor={Vladimir Kaganer},
 pdftex,bookmarksnumbered=true,colorlinks,citecolor=blue,linkcolor=blue,urlcolor=blue}

\makeatletter
\usepackage{newtxtext}

\makeatother

\begin{document}
 
\begin{frontmatter}
\title{Dislocation correlations in GaN epitaxial films revealed by EBSD and
XRD}
\author{Vladimir M. Kaganer\corref{cor1}}
\cortext[cor1]{Corresponding author.}
\ead{kaganer@pdi-berlin.de}
\author{Domenik Spallek}
\author{Philipp John}
\author{Oliver Brandt}
\author{Jonas Lähnemann}
\address{Paul-Drude-Institut für Festkörperelektronik, Hausvogteiplatz 5--7,
10117 Berlin, Germany}
\begin{abstract}
Correlations between dislocations in crystals reduce the elastic energy
via screening of the strain by the surrounding dislocations. We study
the correlations of threading dislocations in GaN epitaxial films
with dislocation densities of $5\times10^{8}$\,cm$^{-2}$ and $1.8\times10^{10}$\,cm$^{-2}$
by X-ray diffraction (XRD) in reciprocal space and by high-resolution
electron backscatter diffraction (HR-EBSD) in real space, where the
strain is derived from a cross-correlation analysis of the Kikuchi
patterns. The measured XRD curves and HR-EBSD strain and rotation
maps are compared with Monte Carlo simulations within one and the
same model for the dislocation distributions. The screening of the
dislocation strains is modeled by creating pairs of dislocations with
opposite Burgers vectors, with the mean distance between dislocations
in a pair equal to the screening distance. The pairs overlap and cannot
be distinguished as separate dipoles. The HR-EBSD-measured autocorrelation
functions of the strain and rotation components follow the expected
logarithmic law for distances smaller than the screening distances
and become zero for larger distances, which is confirmed by the Monte
Carlo simulations. The kink in the plot of the autocorrelation function
allows a robust and accurate determination of the screening distance
without making any simulation or fit. Screening distances of 2 \textmu m
and 0.3 \textmu m are obtained for the samples with low and high dislocation
densities, respectively. The dislocation strain is thus screened by
only 4 neighboring dislocations. In addition, an anisotropic resolution
of the HR-EBSD measurements is observed and quantified. In this version,
an error in the processing of the HR-EBSD maps of the Si wafer is specified.
\end{abstract}
\end{frontmatter}

\section{Introduction}

It is well established that the elastic energy of a single straight
dislocation per unit length of the dislocation line $E\propto(\mu b^{2}/4\pi)\ln(R/r_{c})$
diverges as the lateral crystal size $R$ is increased to infinity.
Here $\mu$ is the shear modulus, $b$ is the length of the Burgers
vector, and $r_{c}$ is the dislocation core radius ($r_{c}\simeq2.6b$)
\citep{HullBacon11ch44}. Hence, the elastic energy density of a crystal
containing a finite density of random \emph{uncorrelated} dislocations
tends to infinity as the crystal size increases. The energy remains
finite when dislocations are correlated, such that the long-range
strain field of a dislocation is screened by surrounding dislocations.
Then, $R$ is the screening distance, rather than the crystal size.

\begin{figure*}
\centering \includegraphics[width=1\textwidth]{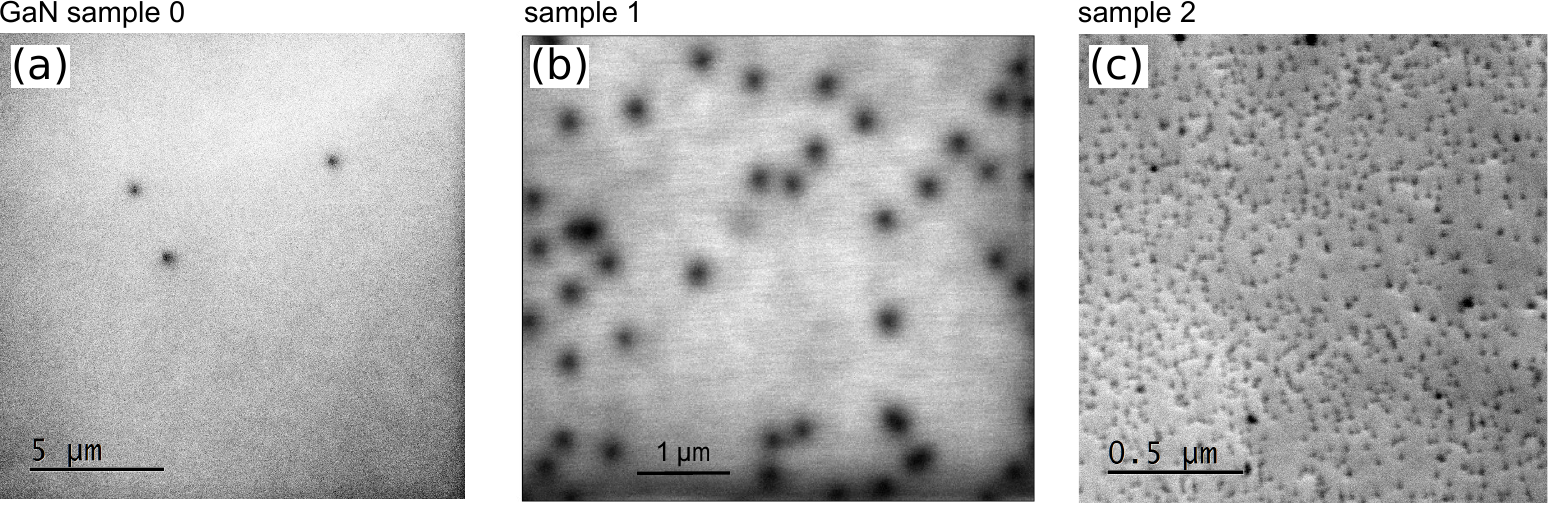}

\caption{(a,b) Panchromatic cathodoluminescence intensity maps of GaN samples
0 and 1 and (c) scanning electron micrograph of sample 2.}

\label{fig:Samples} 
\end{figure*}

Both dislocation density and the screening distance affect the intensity
distributions in X-ray diffraction (XRD). The XRD line profile analysis
\citep{scardi04,gubicza14} is an established method in powder diffraction
for determining the dislocation density as well as the screening distance,
and hence the elastic energy stored in a crystal, and changes to these
quantities during plastic deformation. The calculation of the diffracted
intensity requires a model for the dislocation arrangement. Wilkens
proposed \citep{wilkens69,wilkens70nbs} a ``restrictedly random
dislocation distribution'', where the crystal is subdivided into
cells, with each cell containing the same number of dislocations.
The numbers of dislocations with opposite Burgers vectors in a cell
are equal so that the total Burgers vector of a cell is zero, resulting
in the total strain from the dislocations in a cell decaying with
the distance faster than the strain of an individual dislocation (in
Wilkens' model, the strain is assumed to be restricted within the
cell). The cell size $R$ then plays the role of the screening distance
for dislocation strains. The strain probability density distribution
calculated for screw dislocations in the Wilkens model \citep{wilkens70nbs}
is used in computer programs for XRD line profile analysis \citep{ribarik01,scardi02}.

The XRD intensity distribution can be calculated for any given dislocation
distribution model by means of Monte Carlo integration \citep{kaganer09prbMC,kaganer09prbGaN}.
A Monte Carlo simulation of the XRD profiles that implements the Wilkens
model literally leads to artifacts when a cell contains only one or
a few dislocation pairs \citep{kaganer14acta}. An alternative model
for dislocation strain screening is to consider random pairs of dislocations
with opposite Burgers vectors, with $R$ being the mean distance between
dislocations in a pair \citep{kaganer10acta}. When $R$ exceeds the
mean distance between all dislocations $\varrho^{-1/2}$, the pairs
overlap and cannot be distinguished as individual dipoles. Below,
we only deal with such overlapping pairs and do not consider the case
of dislocation dipoles with the dipole widths small compared to the
separation between dipoles.

In laboratory XRD studies of dislocations, the diffracted intensity
is collected from the entire sample and integrated over the diffraction
vector orientation in powder diffraction or the diffracted wave direction
in a double-crystal diffraction setup for single crystals. The XRD
line profile analysis is based on fitting of the experimental curves
involving two parameters of the dislocation distribution, the dislocation
density $\varrho$ and the screening distance $R$. In many XRD studies,
the dislocation density $\varrho$ is of primary interest, while the
screening distance $R$ is an auxiliary fit parameter required for
accurately determining the dislocation density. However, determination
of the elastic energy stored in a crystal requires an accurate determination
of $R$ \citep{borbely23}. The screening distance $R$ is accessed
as a transition from the Gaussian central part of the diffraction
line to its tails, which show an intensity decay that is proportional
to $q^{-3}$. These tails originate from diffraction in highly disturbed
regions close to dislocation lines and their intensity is independent
of dislocation correlations.

The screening distance $R$ can be accessed directly if spatial strain
maps are measured. $R$ can then be obtained from the strain-strain
correlation function \citep{richter22}. In that work, all components
of the strain and rotation tensors have been measured on the example
of threading dislocations in (In,Ga)N films by the scanning XRD microscopy
technique \citep{chahine14}, using synchrotron X-rays. Mapping in
three non-coplanar reflections was performed to obtain all nine components
of the strain and rotation tensors. Since the dislocation strain depends
on the distance from the dislocation line $r$ by the universal law
$\varepsilon\propto r^{-1}$, the correlation function decreases proportionally
to the logarithm of the distance for distances smaller than $R$,
and becomes zero at larger distances. Plotting the correlation function
against the logarithm of the distance yields a hook-shaped curve.
The kink in this curve provides the screening distance without the
need for any fitting or assumptions about the dislocation correlations.

A variety of lattice-sensitive microscopy techniques for mapping strain
and rotation in crystals is nowadays available \citep{gammer2019}.
A number of synchrotron-based XRD techniques, such as X-ray Laue microdiffraction
\citep{larson2002,ice2009}, dark-field X-ray microscopy \citep{simons15},
scanning XRD microscopy \citep{chahine14}, Bragg coherent diffraction
imaging \citep{pfeifer2006}, or Bragg X-ray ptychography \citep{takahashi2013},
provide access to the full strain tensor. The correlations of strain
fields have been studied to characterize dislocation distributions
by scanning XRD microscopy \citep{richter22} and dark-field X-ray
microscopy \citep{zelenika2024}.

While these XRD techniques are restricted to synchrotron radiation
sources, high-resolution electron back scattering diffraction (HR-EBSD)
is a widely available laboratory technique that provides spatial maps
of all components of the strain and rotation tensors near the crystal
surface. The strains and rotations are obtained from shifts of the
Kikuchi bands \citep{venables73ebsd,schwartz09,wilkinson12}. The
strain sensitivity of the method was enhanced to $10^{-4}$ with the
invention of the a cross-correlation-based analysis of EBSD patterns
\citep{wilkinson06mse,wilkinson06ultra}. Strain and rotation maps,
as well as the strain probability density distributions, have been
reported in several studies \citep{wilkinson14,vilalta17,kalacska17}.
The maps obtained in the present study are similar to those reported
in the previous studies. However, the primary aim of our work is to
obtain and analyze the spatial correlations of the strains, a subject
not addressed in these previous studies. We demonstrate that the strain-strain
correlations calculated from the maps clearly reveal the dislocation
strain screening by surrounding dislocations and provide the screening
distances $R$.

The analysis of dislocations in GaN and other group III nitrides is
of primary interest for optoelectronic applications of these materials
and, at the same time, can serve as a touchstone for various methods
of evaluating dislocation arrays. GaN epitaxial films contain threading
dislocations oriented parallel to each other and perpendicular to
the film surface. These well-defined dislocation arrays enable detailed
modeling of both XRD curves and HR-EBSD maps and their quantitative
comparison. A respective comparison of different methods for dislocation
density determination in metals \citep{gallet23} is much more complicated
since it requires an account of several slip systems, grain boundaries,
etc.

We study GaN(0001) epitaxial films with dislocation densities differing
by orders of magnitude. We compare a film grown by metal-organic chemical
vapor deposition with a threading dislocation density of $5\times10^{8}$\,cm$^{-2}$
and a film grown by molecular beam epitaxy with a dislocation density
of $1.8\times10^{10}$\,cm$^{-2}$. A free standing GaN film with
a dislocation density as low as $6\times10^{5}$\,cm$^{-2}$ and
a perfect Si crystal are used as references.

We carry out both laboratory XRD and HR-EBSD measurements on the same
samples and also perform Monte Carlo simulations of the XRD line profiles
and HR-EBSD strain maps, using one and the same model of dislocation
distributions for both experimental techniques and the same computing
program. Comparing the measured and Monte-Carlo-modeled strain maps
in HR-EBSD provides us with two further findings in addition to the
determination of the screening distance for the dislocation strain.
First, we find that the resolution of HR-EBSD is highly anisotropic,
with significantly worse resolution in the direction of the inclination
of the incident electron beam. We obtain a quantitative estimate of
the resolution by comparing the strain-strain correlations in two
orthogonal directions. Secondly, we find that the experimental HR-EBSD
maps show, as uncorrelated random noise, shear strain components that
should be zero at a strain-free surface. The widths of their distributions
can be used as an estimate of the accuracy of the HR-EBSD measurements.

\section{Experiment and Monte Carlo simulations}

\subsection{Samples}

\label{subsec:Samples}

We study three GaN(0001) samples with different densities of threading
dislocations (see Fig.\,\ref{fig:Samples}). In all samples, the
dislocations are straight lines running from the substrate to the
film surface, along the surface normal. Hence, threading dislocations
provide well defined arrays in single crystal samples suitable for
both XRD and HR-EBSD studies. The investigated samples were characterized
in our previous works \citep{kaganer15jpd,trilogy22_II,trilogy_22_III}.

As a reference, we use a 350\,\textmu m-thick free-standing GaN (0001)
film grown by hydride vapor phase epitaxy. We refer to it as sample
0. The dislocation density in this film is estimated from cathodoluminescence
images to be as low as $6\times10^{5}$\,cm$^{-2}$. Since this sample
produced an unexpectedly broad strain distribution in the HR-EBSD
measurements described in Sec.\,\ref{subsec:EBSD} below, it was
rinsed with isopropanol and ethanol and then cleaned with oxygen plasma
to remove any contaminants and produce a uniformly oxidized surface
layer. Etching of the surface with heated KOH removed this oxidized
layer and left a smooth surface with a root-mean-square roughness
of 0.31\,nm. However, this cleaning step had very little effect on
the HR-EBSD maps.

Sample 1 is a 5.6\,\textmu m-thick GaN(0001) film on an Al$_{2}$O$_{3}$(0001)
substrate. A 1.3-\textmu m-thick GaN layer is fabricated by plasma-assisted
molecular beam epitaxy on top of a 4.3-\textmu m thick GaN(0001) template,
which in turn is grown by metal-organic chemical vapor deposition
on an Al$_{2}$O$_{3}$(0001) substrate. As this sample has been grown
for our former study \citep{trilogy22_II,trilogy_22_III}, an additional
3-nm-thick (In,Ga)N single quantum well is buried at about 650 nm
below the surface, which however is not relevant for the present work.
Threading dislocations, which are in the focus of the present study,
are inherited from the template. Figure \ref{fig:Samples}(b) presents
a cathodoluminescence (CL) intensity image of this sample. Dislocations
are seen as dark spots since they act as centers of nonradiative recombination
of excitons \citep{rosner97,sugahara98,speck99,pauc06,kaganer18apl}.
The density of the spots is $5\times10^{8}$\,cm$^{-2}$. We note
that the counting of the dark spots on CL images may underestimate
the density of threading dislocations \citep{khoury13}.

Sample 2 has been grown for another study of ours \citep{kaganer15jpd}.
A 2.5\,\textmu m-thick GaN(0001) film is grown by plasma-assisted
molecular beam epitaxy on a 6H-SiC(0001) substrate. A cathodoluminescence
image of this sample shows overlapping dark spots because of the large
dislocation density, which does not allow an accurate determination
of the dislocation density. Figure \ref{fig:Samples}(c) presents
a scanning electron micrograph of this sample. It shows dark spots
with the density of $1.8\times10^{10}$\,cm$^{-2}$. These spots
have been identified as pits at dislocation outcrops by comparing
a similar electron micrograph of sample 1 with the CL map taken from
the same surface area \citep{trilogy_22_III}. For the sample with
a low dislocation density, a one-to-one correspondence of the pits
with the dark spots in the CL image has been observed.

Additionally, a reference HR-EBSD measurement is obtained on a Si(001)
wafer.

\subsection{X-ray diffraction}

The XRD measurements were carried out with CuK$\alpha_{1}$ radiation
using a Panalytical X'Pert diffractometer equipped with a two-bounce
Ge(220) hybrid monochromator. The measurements were performed using
the skew diffraction geometry \citep{srikant97,sun02,kaganer05GaN}
in a double-crystal setup.

\subsection{Electron backscatter diffraction}

\label{subsec:EBSD-method}

Scanning electron microscopy (SEM) and CL imaging, as well as the
HR-EBSD measurements were carried out in a Zeiss Ultra 55 scanning
electron microscope. For the HR-EBSD measurements, the microscope
was operated at an acceleration voltage of 15\,kV with a beam current
of 6\,nA. To record the Kikuchi patterns, the stage with the mounted
sample was tilted by 70$^{\circ}$ with respect to the incident beam
direction towards an EDAX Hikari Super EBSD detector.

The patterns were recorded with the highest available resolution of
$470\times470$ pixels and background-corrected using a reference
pattern. HR-EBSD maps were recorded over either $2\times2$\,\textmu m$^{2}$
or $10\times10$\,\textmu m$^{2}$ areas with step sizes of 20\,nm
and 50\,nm, respectively, at an exposure time of 100\,ms. Measurements
with these settings took approximately 20\,min for the smaller maps
and 80\,min for the bigger maps to record. Sample drift was estimated
by comparing SEM images recorded before and after the measurement
and were smaller than 18\%. Its effect is quantified and discussed
at the end of Sec.\,\ref{subsec:EBSD-MC}, as the discussion is based
on the results of Monte Carlo simulations. We find that the effect
of drift can be neglected.

Cross-correlation analysis of the Kikuchi patterns\footnote{Early studies distinguish Kikuchi patterns formed by electron diffraction
in transmission electron microscopy, and pseudo-Kikuchi patterns due
to diffraction of backscattered electrons in SEM \citep{wolf69}.
Nowadays, the term Kikuchi pattern is used for both of them.} \citep{wilkinson09} was carried out with the software CrossCourt
4.5.3.6 from BLG Vantage. Examples of Kikuchi patterns from all four
samples are presented in Fig.\,SM1 in the Supplementary Material
(SM). The software defaults were chosen for 20 regions of interest
(squared subsets of the Kikuchi pattern). A reference pattern was
chosen as the pixel with the lowest kernel average misorientation
in the initial indexation of the EBSD map. The Kikuchi pattern of
each region was compared to the same area of a reference pattern to
identify tiny shifts and rotations of the pattern. The cross-correlation
method enables these changes to be determined with a sub-pixel resolution.
From changes in different parts of the Kikuchi pattern, the strain
and rotation tensor components are obtained using the elastic moduli
of GaN \citep{polian96}. Applying this analysis to every point of
an EBSD map enables the changes in strain and crystal rotation to
be mapped.

\subsection{Monte Carlo simulations}

\label{subsec:MonteCarlo}

Monte Carlo simulations of both XRD curves and HR-EBSD maps are performed
in the framework of one and the same model of dislocation arrangements
and dislocation strains using the computing programs that were developed
earlier for XRD \citep{kaganer15jpd,kaganer24} and adapted here for
HR-EBSD. We include in the simulations all types of perfect threading
dislocations that are present in wurtzite GaN epitaxial layers \citep{metzger98,moram09jap,moram09,bennett10}.
The displacement fields of \emph{a}-type edge dislocations with Burgers
vectors $1/3\left\langle 11\bar{2}0\right\rangle $, \emph{c}-type
screw dislocations with Burgers vectors $\left\langle 0001\right\rangle $,
and \emph{a+c}-type mixed dislocations with Burgers vectors $1/3\left\langle 11\bar{2}3\right\rangle $
are calculated. Crystallographically equivalent Burgers vectors are
taken with equal probabilities.

The screening of the dislocation strains by surrounding dislocations
is provided by generating dislocations in pairs with opposite Burgers
vectors. The positions of the pairs are random and uncorrelated. The
distances between dislocations in the pairs are taken from a lognormal
distribution with a mean value of $R$ (hereafter referred to as the
screening distance) and a standard deviation of $R/2$. The simulations
are insensitive to the standard deviation: a unimodal distribution
(all dislocation pairs have the same distance between dislocations)
gives the same result. The XRD intensity $I(q)$ is calculated as
a probability density of the distortion component $q=\hat{\mathbf{K}}{}^{\mathrm{out}}\cdot\nabla(\mathbf{Q}\cdot\mathbf{U})$,
where $\mathbf{U}(\mathbf{r})$ is the total displacement due to all
dislocations, $\mathbf{Q}$ is the diffraction vector, and $\hat{\mathbf{K}}{}^{\mathrm{out}}$
is the unit vector in the direction of the diffracted beam \citep{kaganer15jpd,kaganer24}.
It is worth noting that the direction $\hat{\mathbf{K}}{}^{\mathrm{out}}$
emerges in the diffraction from a single crystal as a consequence
of the intensity integration performed by a widely open detector over
the plane in reciprocal space perpendicular to $\mathbf{K}^{\mathrm{out}}$.
In powder diffraction, the respective intensity integration from randomly
oriented grains is performed over the plane perpendicular to the diffraction
vector. In this case, $\hat{\mathbf{K}}{}^{\mathrm{out}}$ is replaced
with the unit vector in the direction of $\mathbf{Q}$ \citep{kaganer05GaN}.

The HR-EBSD maps are simulated by calculating all components of the
strain tensor $\varepsilon_{ij}=(\partial U_{i}/\partial x_{j}+\partial U_{j}/\partial x_{i})/2$
and the rotation tensor $\omega_{ij}=(\partial U_{i}/\partial x_{j}-\partial U_{j}/\partial x_{i})/2$,
using the same displacement vector $\mathbf{U}(\mathbf{r})$. For
the XRD simulations, it is sufficient to use the displacement field
of a straight dislocation in an infinite crystal, as the X-ray intensity
is collected over the entire film thickness. The displacement for
a dislocation along the $\left\langle 0001\right\rangle $ direction
in a hexagonal crystal is employed. It coincides with the isotropic
solution with the Poisson ratio taken to be $\nu_{h}=c_{12}/(c_{11}+c_{12})$
(\citep{belov92}, Sec.\,2.5). Here $c_{ij}$ are the elastic moduli.
Using their values for GaN \citep{polian96}, we obtain $\nu_{h}=0.27$.
For the HR-EBSD simulations, the elastic strain relaxation at the
free surface is taken into account. We have compared the isotropic
solution (\citep{lothe92}, Sec.\,6.3) with a more comprehensive
anisotropic solution for a dislocation along the sixfold axis (\citep{belov92},
Sec.\,2.5). The difference is negligible compared to the statistical
errors in the Monte Carlo simulations presented below.

\section{Results}

\subsection{X-ray diffraction}

\begin{figure*}
\centering \includegraphics[width=1\textwidth]{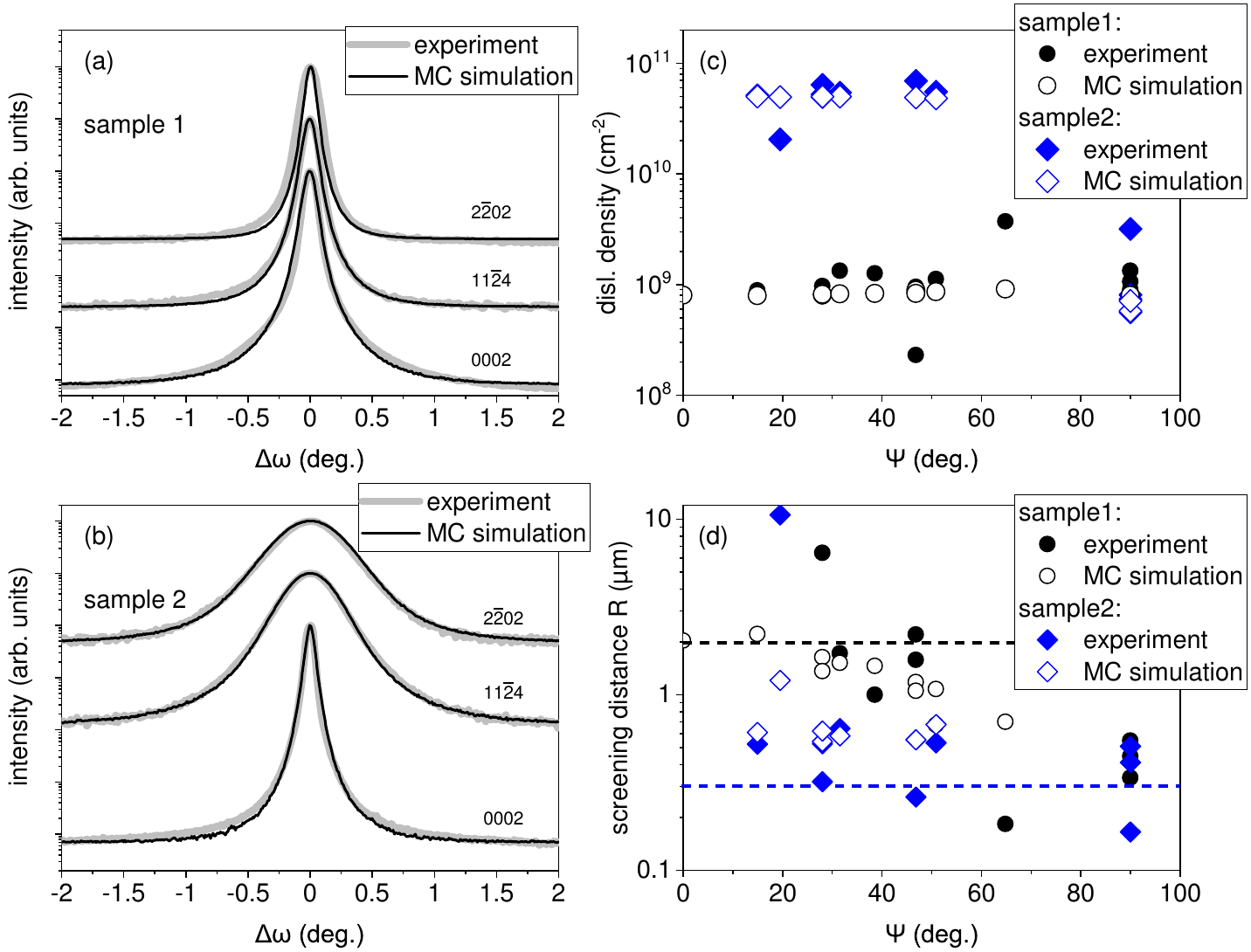}

\caption{(a,b) X-ray diffraction curves of samples 1 and 2 (thick gray lines)
and their Monte Carlo simulations (thin black lines). (c,d) Dislocation
density and the screening distance $R$ obtained in the fits of the
experimental (full symbols) and the simulated (open symbols) curves.
The fits for different reflections are made independently and presented
as a function of the angle $\Psi$ between diffraction vector and
surface. The parameters for sample 1 are presented by black symbols,
and for sample 2 by blue symbols. Dashed lines in (d) mark the values
of screening distances obtained in the analysis of EBSD maps. }

\label{fig:XRD} 
\end{figure*}

Figures \ref{fig:XRD}(a) and \ref{fig:XRD}(b) show by thick gray
lines measured XRD curves for several reflections from samples 1 and
2, respectively. A symmetric Bragg reflection $0002$ and two asymmetric
ones,\footnote{We refer to these reflections as asymmetric following the common tradition.
In fact, the diffraction vectors of these reflections are inclined
with respect to the surface normal, and these reflections are asymmetric
in coplanar diffraction. In skew geometry, however, these reflections
are symmetric (the incident and the diffracted waves make equal angles
with the crystal surface) and non-coplanar (the scattering plane does
not contain the surface normal) \citep{kaganer05GaN}.} $11\bar{2}4$ and $2\bar{2}02$, are presented. Some qualitative
conclusions can be drawn from a direct comparison of the curves. The
diffraction vector of the symmetric reflection is directed along the
lines of threading dislocations. Consequently, this reflection is
only sensitive to screw dislocations and screw components of mixed
dislocations, which produce displacements in the direction of the
dislocation lines. The widths of the symmetric and asymmetric reflections
from sample 1 are closely comparable, indicating similar densities
of screw and edge dislocations. In contrast, the symmetric reflection
from sample 2 is notably narrower than the asymmetric reflections
from this sample, which is due to a low density of screw dislocations
in comparison with the edge ones. Comparing diffraction curves from
the two samples, one can see that the density of screw dislocations
in sample 2 is smaller than in sample 1, while the density of edge
dislocations is notably larger.

The thin black lines in Figs.\,\ref{fig:XRD}(a,b) represent Monte
Carlo simulations of the respective diffraction curves. First, we
simulate the symmetric Bragg reflections of sample 1 with screw dislocations.
Agreement with the experimental diffraction profiles is obtained with
a density of $8\times10^{8}$\,cm$^{-2}$ and the screening distance
of $R=1$\,\textmu m. Next, we turn to asymmetric reflections and
include the edge components of the Burgers vectors in the calculations.
Since screw dislocations are very uncommon in GaN(0001) films grown
by metal-organic chemical vapor deposition \citep{metzger98,moram09jap,moram09,bennett10},
we attribute the density of screw dislocations obtained in the symmetric
reflections to the screw components of mixed \emph{a+c}-type dislocations.
The calculation of the diffraction profiles of asymmetric reflections
with mixed dislocations agrees with the experimental diffraction curves,
meaning that we do not need to add extra edge dislocations. We therefore
conclude that sample 1 contains only mixed dislocations.

For sample 2 grown by molecular beam epitaxy, we also first simulate
the symmetric Bragg reflections with screw dislocations. This revealed
a screw dislocation density of $5\times10^{8}$\,cm$^{-2}$ and a
screening distance of $R=0.7$\,\textmu m. These dislocations provide
only a small contribution to the widths of the asymmetric reflections.
To simulate these reflections, we add edge dislocations with a density
of $5\times10^{10}$\,cm$^{-2}$, two orders of magnitude higher
than the density of screw dislocations, with the screening distance
of $R=0.3$\,\textmu m.

We fit both the measured and the Monte-Carlo-simulated XRD curves
in the same way, following Ref.\,\citep{kaganer05GaN}. The symmetric
reflections are fitted first to get the densities of screw dislocations,
and their contributions are taken into account in the fits of the
asymmetric reflections, which give the densities of the edge dislocations.
Each reflection is fitted independently from the others. The two essential
fitting parameters are the dislocation density $\varrho$ and the
screening distance $R$. Figures \ref{fig:XRD}(c,d) present by full
symbols the values of $\varrho$ and $R$ obtained in the fits of
the measured XRD curves, while open symbols are the results of the
corresponding fits of the Monte-Carlo-simulated curves. The data are
plotted as a function of the angle $\Psi$ between the diffraction
vector and the surface. Thus, $\Psi=0$ corresponds to an in-plane
reflection and $\Psi=90^{\circ}$ to a symmetric Bragg reflection.
Plotting the fit parameters as a function of $\Psi$ is just a convenient
way of presenting the data: ideally, no variation of the parameters
is expected.

The error bars representing the possible errors in determining the
fit parameters from a fitted curve of a given reflection are smaller
than the symbol size in Figs.~\ref{fig:XRD}(c,d) and can therefore
be neglected. The scattering of the data points obtained from fitting
the experimental curves in different reflections can have two sources,
the accuracy of the fit function proposed in Ref.\,\citep{kaganer05GaN}
and the presence of additional sources of strain, particularly a contribution
from dislocations at the interface between the substrate and the relaxed
epitaxial films. Our Monte Carlo simulations enable us to estimate
the effect of the former by comparing the input parameters of the
simulation with the output parameters of the fit.

For sample 1, the input parameters for the Monte Carlo simulation
were the dislocation density $\varrho=8\times10^{8}$~cm$^{-2}$
of mixed dislocations and the average width of the dislocation pairs
$R=1$~\textmu m. Fitting the Monte Carlo simulated curves gives
a dislocation density $\varrho=(8.3\pm0.4)\times10^{8}$~cm$^{-2}$
and an average width of the dislocation pairs $R=1.4\pm0.45$~\textmu m.
The errors represent the scattering between data points of different
reflections in Figs.~\ref{fig:XRD}(c,d). For sample 2, the input
parameters for the Monte Carlo simulations were the density $\varrho=5\times10^{10}$~cm$^{-2}$
of edge dislocations and $R=0.3$~\textmu m. The fit of the simulated
curves gives $\varrho=(5\pm0.06)\times10^{10}$~cm$^{-2}$ and $R=(0.5\pm0.05)$~\textmu m.
These results show that the fits following Ref.\,\citep{kaganer05GaN}
quite accurately reproduce the dislocation density, but overestimate
the screening distance by about 50\%. This is not surprising since
the fit formula for XRD curves contains prefactors for the dislocation
density and for the screening distance. The prefactor for the dislocation
density, also referred to as the ``contrast factor'' in powder diffraction,
depends on the relative orientations of all the vectors involved in
the problem (the diffraction vector, the Burgers vector, and the dislocation
line direction). It can be accurately calculated in powder diffraction
\citep{martinez09} and in diffraction from single crystals \citep{kaganer05GaN}.
A respective prefactor for the screening distance has only been calculated
analytically in the case of screw dislocations in the Wilkens' model
\citep{wilkens70nbs}. It can only be estimated for other cases, which
limits the accuracy of determining the screening distance from the
XRD profiles \citep{borbely23}.

Fitting the experimental curves in different reflections yields for
sample 1 a density $\varrho=(1\pm0.3)\times10^{8}$~cm$^{-2}$ of
mixed dislocations and $R=1\pm0.7$~\textmu m, and for sample 2 a
density $\varrho=(5.8\pm0.7)\times10^{10}$~cm$^{-2}$ of edge dislocations
and $R=0.46\pm0.14$~\textmu m. These dislocation densities are larger
than the densities of the spots in the cathodoluminescence or scanning
electron microscopy images in Fig.\,\ref{fig:Samples} by factors
of 2 for sample 1 and 3 for sample 2. We note that similar overestimates
of the dislocation density in comparison with transmission electron
microscopy results were obtained in previous studies \citep{kaganer05GaN,kopp13jac,kopp14jap}.

The HR-EBSD study presented in the next section complements the reciprocal-space
analysis by XRD with a real-space observation of the dislocation correlations.
In this way, the screening distance $R$ can be determined more accurately
and independent of the dislocation density determination. The values
of $R$ obtained from the HR-EBSD maps are marked by dashed lines
in Fig.\,\ref{fig:XRD}(d) for comparison.

\subsection{Electron backscatter diffraction: experiment}

\label{subsec:EBSD}

\begin{figure*}
\centering \includegraphics[width=1\textwidth]{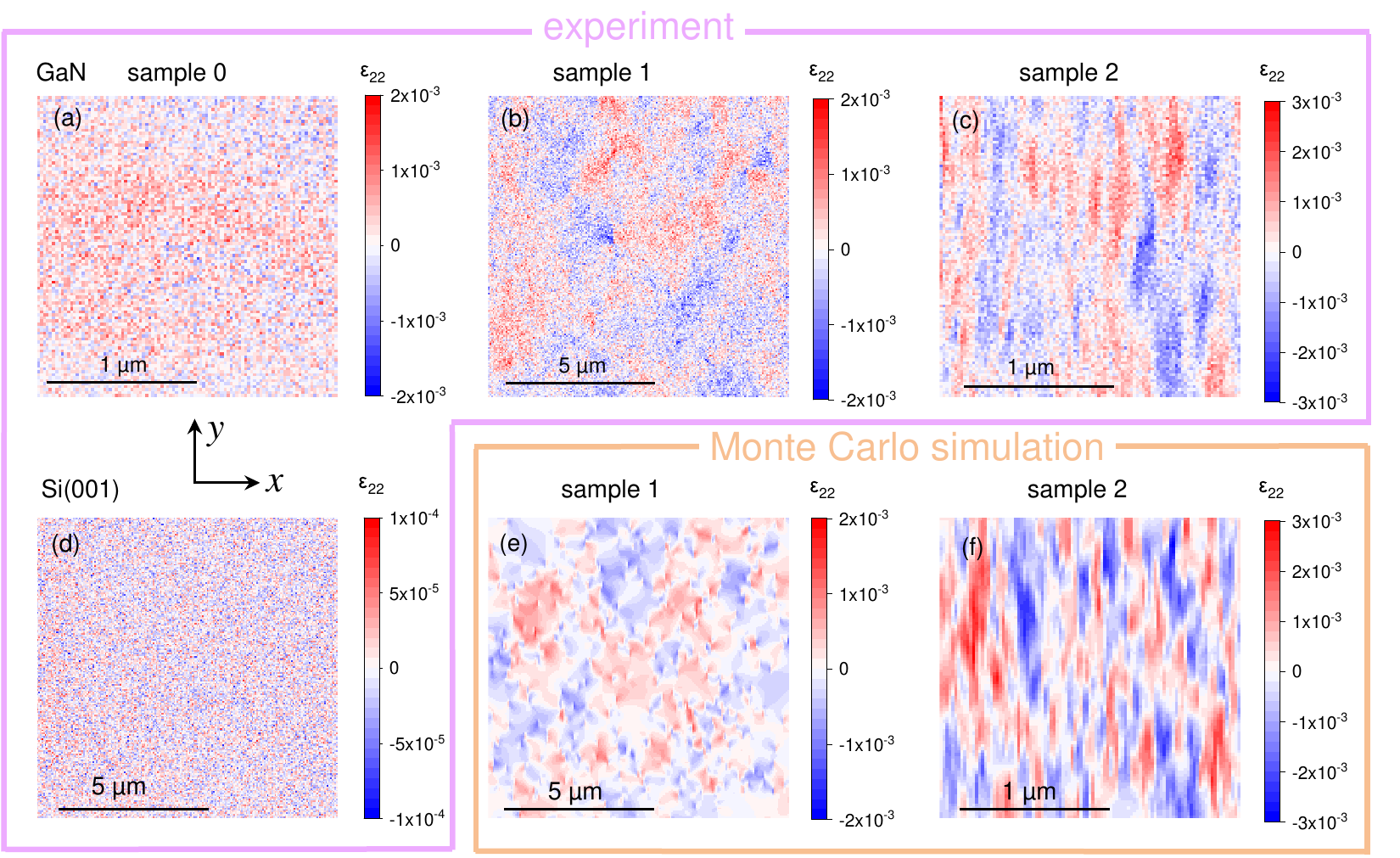}

\caption{HR-EBSD maps of the strain component $\varepsilon_{22}$ (a--c) in
GaN samples 0, 1, and 2 and (d) in a Si(001) wafer, and (e,f) Monte
Carlo simulation of the $\varepsilon_{22}$ maps for samples 1 and
2. Note that the strain obtained for Si(001) is an order of magnitude
lower than in the GaN samples. The $y$ axis is chosen in the direction
of the tilt of the electron beam.}

\label{fig:mapsEps22} 
\end{figure*}

Figures \ref{fig:mapsEps22}(a--c) present maps of the strain component
$\varepsilon_{22}$ of the GaN samples 0, 1, and 2 obtained from the
analysis of the HR-EBSD measurements. Complete sets of maps of the
six components of the strain tensor $\varepsilon_{ij}$ and the three
components of the rotation tensor $\omega_{ij}$ are presented in
the SM, Figs.\,SM2, SM3, and SM4. As the HR-EBSD maps are relative
to an arbitrarily chosen reference value, we set the mean value of
each map to zero. We follow the standard HR-EBSD axes notation: the
sample surface is the $xy$ plane, the $y$ axis is oriented such
that the incident electron beam (inclined by $70^{\circ}$ with respect
to the surface normal) is in the $yz$ plane. The sizes of the maps,
$10\times10$\,\textmu m$^{2}$ for sample 1 and $2\times2$\,\textmu m$^{2}$
for sample 2, are chosen to cover a sufficiently large number of dislocations,
of the order of $10^{3}$, for a statistical analysis.

In a $2\times2$\,\textmu m$^{2}$ map of sample 0, just a single
dislocation can be expected with a probability of 25\%, compared to
500--700 dislocations in the maps of samples 1 and 2. So, sample
0 can be considered as dislocation free on the scale of the HR-EBSD
measurements. Nevertheless, the range of the strain variation in Fig.\,\ref{fig:mapsEps22}(a)
is comparable with that in samples 1 and 2. In contrast to these latter
samples, the strain map does not reveal correlations between strains
in neighboring measurement points, which is quantified below by calculation
of the strain--strain correlations. Since the strain range for sample
0 in Fig.\,\ref{fig:mapsEps22}(a) is an order of magnitude larger
than the expected sensitivity of the HR-EBSD measurements, the same
measurement has been performed on a Si(001) wafer as presented in
Fig.\,\ref{fig:mapsEps22}(d). The map also displays uncorrelated
random noise, but the range of strain variations is of the order of
$10^{-4}$, which is the expected sensitivity of HR-EBSD \citep{wilkinson06mse,wilkinson06ultra}
and an order of magnitude smaller than in all GaN samples under investigation.
Whole sets of maps of all components of the strain and rotation tensors
of Si(001) are presented in Fig.\,SM5 in the SM and show the variations
in the same range of $10^{-4}$.

The map of $\varepsilon_{22}$ from sample 2 in Fig.\,\ref{fig:mapsEps22}(c)
shows a notable difference between the $x$ and $y$ directions, with
all features being extended in the $y$ direction. Figure SM4 shows
a similar anisotropy in the maps of all strain and rotation components
of this sample. We have performed the same measurement with the sample
rotated by 90$^{\circ}$ around its normal and found that the direction
of the extension in the maps is also rotated by 90$^{\circ}$ to the
new $y$ direction. A possible anisotropy of the sample is thus ruled
out and it is established that the features in the maps are extended
along the direction of the tilt of the incident electron beam. We
show below that this extension is explained by a strongly anisotropic
resolution of the HR-EBSD measurements, and quantify this anisotropy.

\begin{figure*}
\centering \includegraphics[width=1\textwidth]{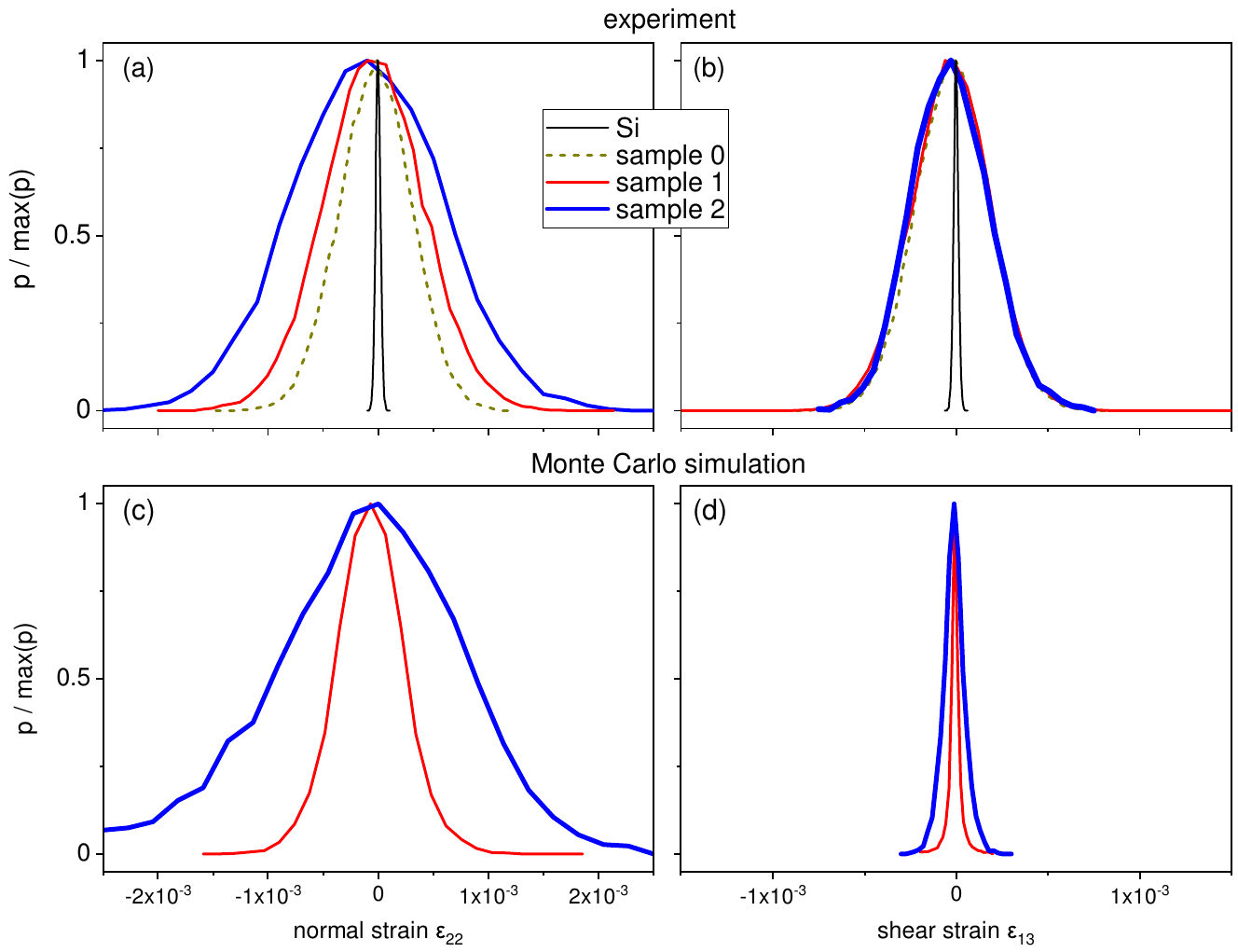}

\caption{Top row: probabilities of (a) normal strain component$\varepsilon_{22}$
and (b) shear strain component $\varepsilon_{13}$ in samples 1 and
2, as well as in the reference dislocation free GaN sample (sample
0) and a silicon wafer, obtained from the maps in Figs.\,\ref{fig:mapsEps22}(a--d).
Bottom row: probabilities of (c) normal strain component$\varepsilon_{22}$
and (d) shear strain component $\varepsilon_{13}$ in the Monte Carlo
simulated maps in Figs.\,\ref{fig:mapsEps22}(e,f) and SM3, SM4.
The probabilities are scaled by their maxima.}

\label{fig:HistogramsStrain} 
\end{figure*}

Figure \ref{fig:HistogramsStrain}(a) presents by red and blue lines
the probability distributions of the strain $\varepsilon_{22}$ obtained
from the maps of samples 1 and 2 in Figs.\,\ref{fig:mapsEps22}(b)
and \ref{fig:mapsEps22}(c). The dashed orange line is a similar distribution
obtained from the map of the dislocation free GaN sample 0 in Fig.\,\ref{fig:mapsEps22}(a),
while the thin black line is the respective distribution from the
Si(001) sample in Fig.\,\ref{fig:mapsEps22}(d)
\footnote{After the paper has been published, we found an error in the 
processing of the HR-EBSD maps of the Si wafer, which affects the strain 
scale of Fig. 3(d) and consequently the widths of the strain distributions 
for Si in Figs. 4(a) and (b). This error was corrected in the 
Corrigendum [74] and added to the Supplementary Material of this version 
as Sec.\ SM 6.}. A comparison of
the widths of the strain distributions from dislocation free GaN and
Si samples points to a material specific broadening for GaN. The possible
origin of this broadening is discussed in Sec.\,\ref{sec:Discussion}
below. The contribution of dislocations in samples 1 and 2 to the
widths of the strain distributions is only moderate.

The shear strains $\varepsilon_{13}$ and $\varepsilon_{23}$ are
of separate interest. Since the information depth of EBSD is small
(less than 20\,nm), the conditions of the stress-free surface $\sigma_{i3}=0\,(i=1,2,3)$,
are satisfied. One of these conditions, $\sigma_{33}=0$, is used
in the processing of the Kikuchi patterns, since only the differences
of the strains $\varepsilon_{11}-\varepsilon_{33}$ and $\varepsilon_{22}-\varepsilon_{33}$
can be found from the positions of the Kikuchi bands and this condition
allows the three normal strain components to be determined separately
\citep{vilalta17}. The other two conditions $\sigma_{i3}=0\,(i=1,2)$
are not used in the processing of the Kikuchi patterns. Since $\sigma_{i3}=2c_{44}\varepsilon_{i3}\,(i=1,2)$
in a hexagonal crystal, the strains $\varepsilon_{13}$ and $\varepsilon_{23}$
must be zero at the free surface. Figure \ref{fig:HistogramsStrain}(b)
shows the probability distributions of $\varepsilon_{13}$ obtained
from the respective HR-EBSD maps of GaN samples 0, 1 and 2 as well
as the silicon wafer. The probability distributions of the other shear
strain component $\varepsilon_{23}$ are about two times narrower
than those of $\varepsilon_{13}$. The maps of these components (shown
in Figs.\,SM2--SM5 of the SM) exhibit uncorrelated random noise,
as do all maps of the GaN sample 0 and the Si wafer. Figure \ref{fig:HistogramsStrain}(b)
shows that all three GaN samples possess the same width of the $\varepsilon_{13}$
distribution, which is 16 times larger than the respective distribution
for Si, and a factor of 2 narrower than the distribution of $\varepsilon_{22}$
from sample 0 in Fig.\,\ref{fig:HistogramsStrain}(a). Hence, the
non-zero shear strains $\varepsilon_{13}$ and $\varepsilon_{23}$
are material specific. We postpone further discussion to Sec.\,\ref{sec:Discussion}.

\begin{figure*}
\centering \includegraphics[width=1\textwidth]{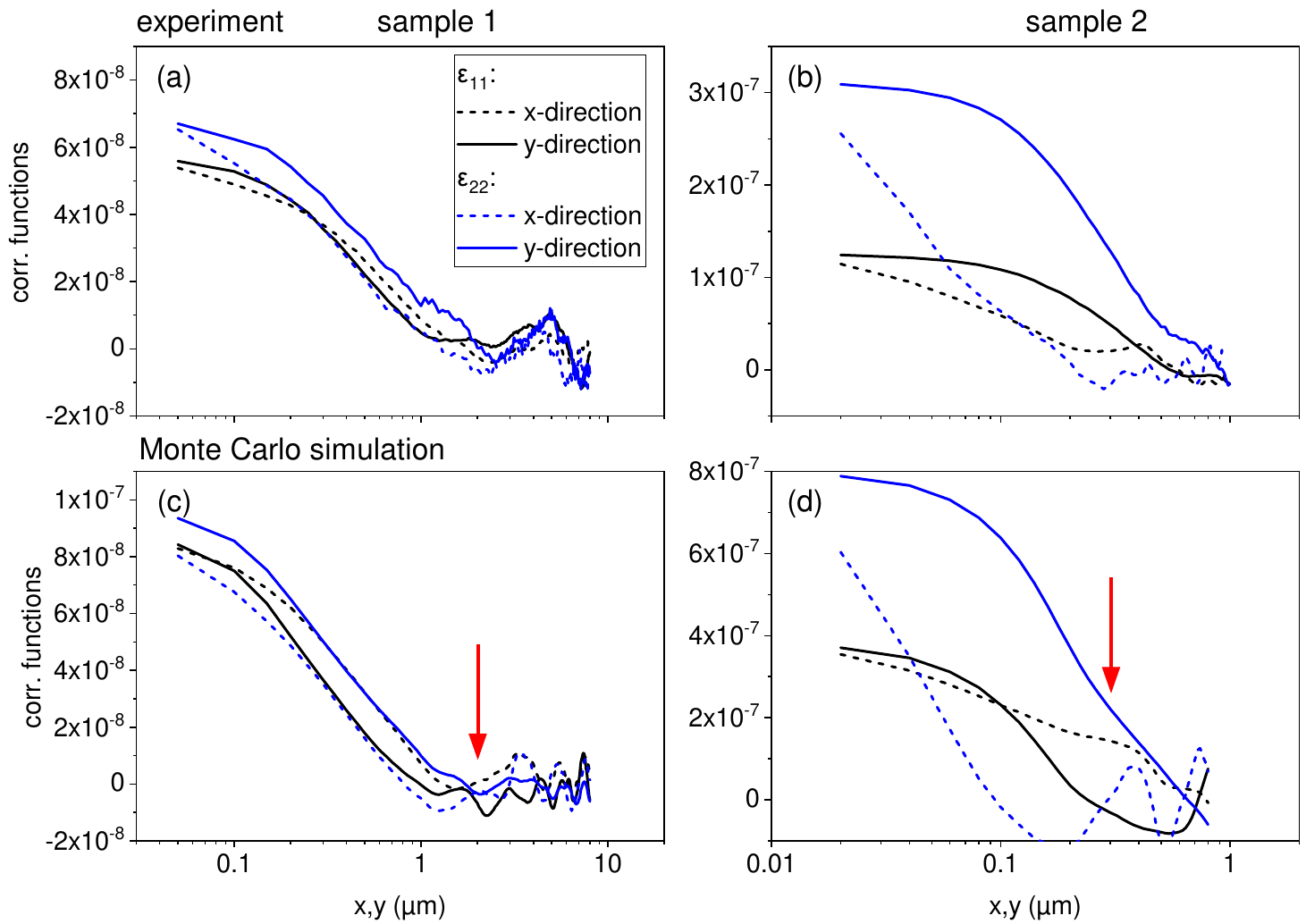}

\caption{Strain--strain correlation functions of (a,b) samples 1 and 2 and
(c,d) their Monte Carlo simulations. The vertical red arrows point
out to the screening distances taken as input of the Monte Carlo simulations
(2~\textmu m and 0.3~\textmu m for samples 1 and 2 respectively).}

\label{fig:CorrFuncXY} 
\end{figure*}

Figures \ref{fig:CorrFuncXY}(a,b) and \ref{fig:CorrFuncStrain}(a,b)
present the strain-strain correlations calculated from the maps of
$\varepsilon_{22}$ in Figs.\,\ref{fig:mapsEps22}(a,b) and from
similar maps of other strain components presented in Figs.\,SM3 and
SM4. We calculate the correlation functions as averages $\left\langle \varepsilon(\mathbf{r}_{1})\varepsilon(\mathbf{r}_{2})\right\rangle $
over the respective maps. Figure \ref{fig:CorrFuncXY} shows correlations
in $x$ and $y$ directions (the difference $\mathbf{r}_{1}-\mathbf{r}_{2}$
is directed either along $x$ or $y$), while Fig.\,\ref{fig:CorrFuncStrain}
shows correlations as a function of the distance $r=\left|\mathbf{r}_{1}-\mathbf{r}_{2}\right|$
averaged over all orientations of $\mathbf{r}_{1}-\mathbf{r}_{2}$.
Since the strain of a single dislocation decays with the distance
as $\varepsilon(\mathbf{r})\propto r^{-1}$, we expect the correlations
to decay as $\ln\left|\mathbf{r}_{1}-\mathbf{r}_{2}\right|$ as long
as $\left|\mathbf{r}_{1}-\mathbf{r}_{2}\right|$ is less than the
distance $R$, at which the dislocation strain is screened by surrounding
dislocations, and to become zero at larger distances. We therefore
plot the correlations in Figs.\,\ref{fig:CorrFuncXY} and \ref{fig:CorrFuncStrain}
as a function of the logarithm of the distance. The expected linear
decay of the correlations in this scale and a kink from the logarithmic
decay to zero correlations are clearly seen in Figs.\,\ref{fig:CorrFuncStrain}(a,b)
and allow the determination of the screening distances: these distances
are 2\,\textmu m for sample 1 and 0.3\,\textmu m for sample 2. This
finding is the main result of the present paper.

Figures \ref{fig:CorrFuncStrain}(a,b) also show a drastic difference
in the correlations between the in-plane strain components $\varepsilon_{11}$,
$\varepsilon_{12}$, $\varepsilon_{22}$ and the shear strains $\varepsilon_{13}$,
$\varepsilon_{23}$. The latter ones are expected to be zero due to
zero stress conditions at the free surface but show distributions
in Fig.\,\ref{fig:HistogramsStrain}(b) comparable in width to the
in-plane strains. Their correlation functions in Figs.\,\ref{fig:CorrFuncStrain}(a,b)
are negligibly small compared to the correlations of the in-plane
strains and quantify the visual impression of the respective maps
in Figs.\,SM3 and SM4 as uncorrelated random noise.

One can see in Figs.\,\ref{fig:CorrFuncXY} and \ref{fig:CorrFuncStrain}
deviations from straight lines (i.e.~from the logarithmic law) at
separations smaller than about 200\,nm: the curves are gradually
flattened. One can also see in Fig.\,\ref{fig:CorrFuncXY}(b) a strong
difference between correlations of $\varepsilon_{11}$ and $\varepsilon_{22}$
and a strong difference between the correlations of $\varepsilon_{22}$
in $x$ and $y$ directions. This plot thus quantifies the anisotropy
of the map of $\varepsilon_{22}$ from sample 2 noted above. The respective
plots for sample 1 in Fig.\,\ref{fig:CorrFuncXY}(a) do not show
such a difference between correlations of $\varepsilon_{11}$ and
$\varepsilon_{22}$ and between the $x$ and $y$ directions because
of the larger step size employed for these maps.

\subsection{Electron backscatter diffraction: Monte Carlo simulations}

\label{subsec:EBSD-MC}

\begin{figure*}
\centering \includegraphics[width=1\textwidth]{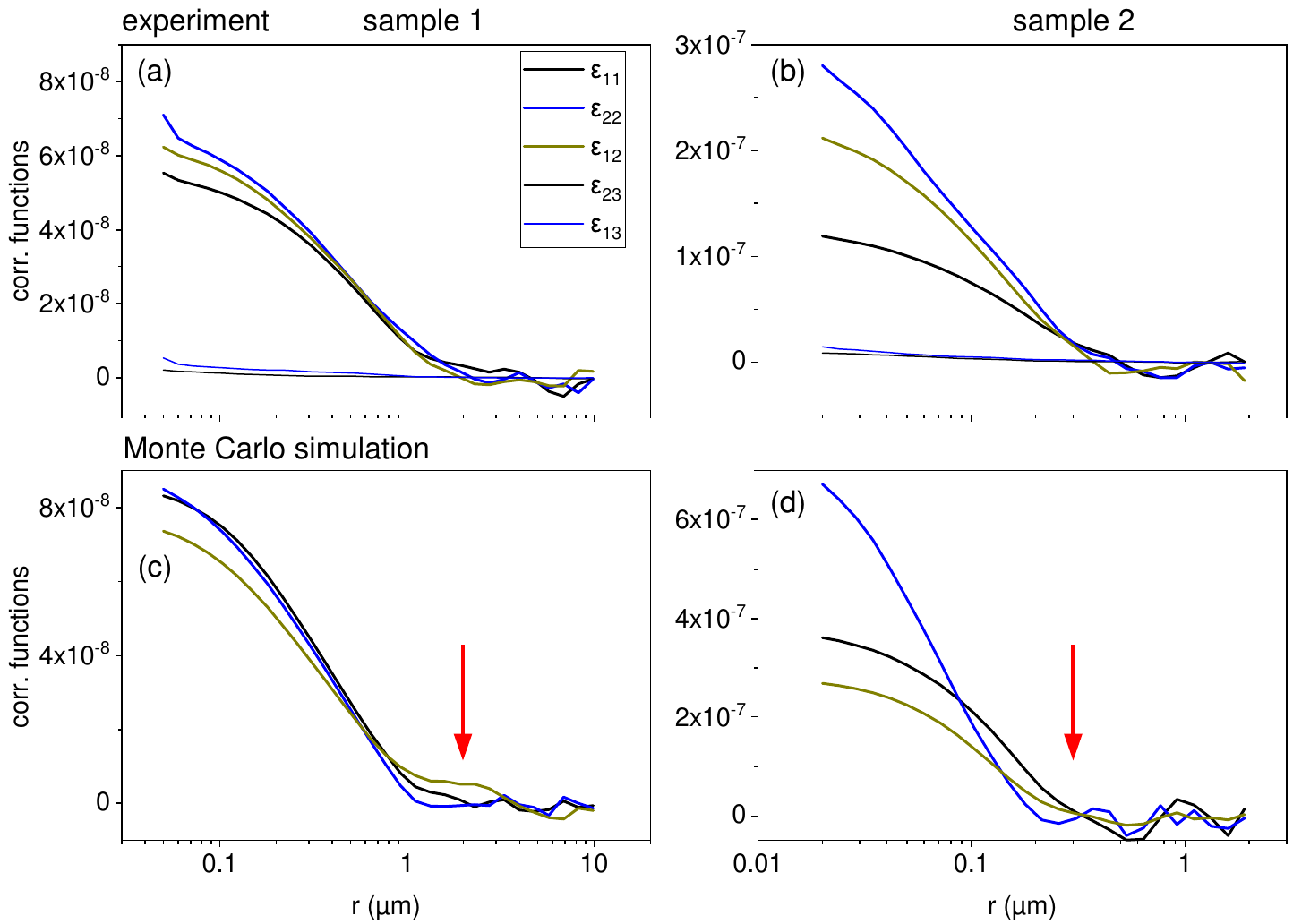}

\caption{Orientation-averaged strain--strain correlation functions of (a,b)
samples 1 and 2 and (c,d) their Monte Carlo simulations. The vertical
red arrows point out to the screening distances taken as input of
the Monte Carlo simulations (2~\textmu m and 0.3~\textmu m for samples
1 and 2 respectively).}

\label{fig:CorrFuncStrain} 
\end{figure*}

We simulate the maps of all strain and rotation components by the
Monte Carlo method, as described in Sec.\,\ref{subsec:MonteCarlo}.
We take the dislocation densities $5\times10^{8}$\,cm$^{-2}$ and
$1.8\times10^{10}$\,cm$^{-2}$ for samples 1 and 2, as determined
in Sec.\,\ref{subsec:Samples}, and the mean distances between dislocations
in the pairs of 2\,\textmu m and 0.3\,\textmu m, respectively, as
derived above from the plots in Fig.\,\ref{fig:CorrFuncStrain}(a,b).
We simulate maps having the same areas and number of pixels as in
the experiment, with the aim to obtain the same level of statistical
errors due to averaging over the same number of dislocations. The
strains are integrated over a 20~nm thick layer at the surface, corresponding
to the information depth of EBSD measurements. The simulated maps
of $\varepsilon_{22}$ in Fig.\,\ref{fig:mapsEps22}(e,f), as well
as the simulated maps of all strain and rotation components shown
in Figs.\,SM6 and SM7 show a good qualitative agreement with the
measured maps.

To make a quantitative comparison, we process the simulated maps to
obtain the correlation functions in the same way as done for the measured
maps. An agreement between simulations and experiment for sample 2
is reached when we introduce a strongly anisotropic resolution to
the simulated maps. Specifically, the calculated strains are averaged
over an area $20\times200$\,nm$^{2}$, extended in the direction
of inclination of the incident electron beam. Such a resolution allows
to reach an agreement between the correlations in $x$ and $y$ directions,
as seen in Figs.\,\ref{fig:CorrFuncXY}(b) and \ref{fig:CorrFuncXY}(d).
The same resolution is imposed for sample 1 but does not give rise
to a similar anisotropy of the maps, because the measurements are
performed with a larger step size, to cover a larger surface area.
As a result, both the experimental and the simulated correlation curves
of sample 1 in Figs.\,\ref{fig:CorrFuncXY}(a) and \ref{fig:CorrFuncXY}(c)
do not show the anisotropy observed for sample 2.

The measured strain distribution of sample 1 in Fig.\,\ref{fig:HistogramsStrain}(a)
is slightly broader than the respective Monte Carlo simulation in
Fig.\,\ref{fig:HistogramsStrain}(c). The additional broadening can
be attributed to the apparent strain observed in sample 0 and discussed
further in Sec.\ \ref{subsec:DiscussResolution}, which is independent
of the dislocation strain.

Figure \ref{fig:HistogramsStrain}(d) presents the probability distributions
of the strain component $\varepsilon_{13}$ obtained from the Monte
Carlo simulated maps of samples 1 and 2 in Figs.\,SM6 and SM7, respectively.
This strain is equal to zero at the surface because of the strain-free
boundary conditions. For edge dislocations, $\varepsilon_{13}$ and
$\varepsilon_{23}$ are also equal to zero in the bulk of the crystal
far from the surface. The non-zero strains in Figs.~\ref{fig:HistogramsStrain}(d)
and SM6, SM7 result from integrating the three-dimensional strain
relaxation field near the surface over the 20~nm-thick surface layer
corresponding to the EBSD information depth. The widths of the probability
distributions of $\varepsilon_{13}$ in the simulated maps in Fig.\ \ref{fig:HistogramsStrain}(d)
are at least 5 times smaller than the respective experimental widths
in Fig.\,\ref{fig:HistogramsStrain}(b). Hence, the strains $\varepsilon_{13}$,
$\varepsilon_{23}$ in the experimental maps are not the strains near
the surface. They look, similarly to all maps of the dislocation-free
sample 0, like uncorrelated random noise and may have the same origin,
which is discussed in Sec.\,\ref{subsec:DiscussResolution}.

The orientation-averaged correlation curves in Fig.\ \ref{fig:CorrFuncStrain}(c,d),
obtained from the Monte Carlo simulated maps, reveal the logarithmic
decay of the strain correlations (straight lines in the coordinates
of the figure) and a kink as a transition to zero correlations at
distances exceeding the screening distance. The positions of the kinks
coincide with the screening distances of 2~\textmu m and 0.3~\textmu m
for samples 1 and 2 respectively, taken as an input of the Monte Carlo
simulations. A gradual increase of correlations is seen at distances
smaller than the resolution of 200\,nm.

The probability distributions and the correlation functions for the
different rotation components, both the measured and the Monte-Carlo
simulated ones, are shown in Figs.\,SM8 and SM9 of the SM. Edge threading
dislocations dominate in sample 2 and give rise to only in-plane rotations
$\omega_{12}$ in the bulk of the sample. However, strain relaxation
at the free surface results in out-of-plane rotations $\omega_{13}$
and $\omega_{23}$. These rotations are smaller than $\omega_{12}$
but larger than the shear strains $\varepsilon_{13}$ and $\varepsilon_{23}$
in Fig.\,\ref{fig:HistogramsStrain}(d), since the out-of-plane rotations
$\omega_{13}$ and $\omega_{23}$ for edge dislocations are not zero
at the surface. The correlation functions of rotations presented in
Fig.\,SM9 show the same logarithmic decay of correlations at distances
smaller than $R$ and the absence of correlations at larger distances.
The characteristic kink is seen at the same distances $R$ as in the
strain correlations of the respective samples.

The mixed dislocations used to model sample 1 give rise to equal widths
of probability distributions of the in-plane and out-of-plane rotations
in the Monte Carlo simulations in Fig.\,SM8(c). The experimental
distribution of the in-plane rotation $\omega_{12}$ in Fig.\,SM8(a)
is broader than the distributions of $\omega_{13}$ and $\omega_{23}$.
In order to achieve agreement, edge dislocations with a density of
$1\times10^{9}$~cm$^{-2}$ would need to be added. However, that
disagrees with the dislocation density obtained from the cathodoluminescence
image in Fig.\,\ref{fig:Samples} and also with the close widths
of the XRD profiles in symmetric and asymmetric Bragg reflections
in Fig.\,\ref{fig:XRD}(a), which point out to mixed dislocations
as a main part of the dislocation ensemble in sample 1. Therefore,
only mixed dislocations are included in the Monte Carlo simulations
of sample 1.

As we noted in Sec.~\ref{subsec:EBSD-method}, the measured EBSD
maps suffer from sample drift, which is estimated to be below 18\%.
While the drift has no effect on the strain probability distributions,
it can potentially affect the correlations of the strain tensor components,
which are the main focus of our study. However, 18\% drift refers
to the relative displacement of the opposite corners of a measured
map. Since we are looking for correlations between points separated
by distances smaller than the screening distance $R$ (the correlations
are absent at larger distances anyway), and since the map sizes are
chosen to be an order magnitude larger than $R$, the effect of drift
on separations smaller than $R$ is negligible. This is quantified
in Fig.\,SM10 in the SM. In this example, we artificially expand
a Monte Carlo-simulated map by 15\% and compare the autocorrelation
functions of the strain maps before and after the stretching. The
difference between the curves can be neglected.

To estimate a possible statistical error in the screening distance
determination, we collect in a single plot the autocorrelation functions
obtained in measurements at different areas of the samples. They are
compared with repeated Monte Carlo simulations performed using the
same parameters, but with different values for the seed of the random
number generator. These plots are compared in Fig.~SM11 of the SM.
Similar scattering of the curves is observed in both the experiment
and the simulations. Therefore, the limited number of dislocations
in a map area (only 500--700) rather than possible sample inhomogeneity
is a main factor limiting the accuracy of the screening length determination.
This could potentially be improved by measuring maps from sample areas
containing larger number of dislocations. From Fig.\,SM11, the accuracy
of our determination of the screening distance is estimated to be
about 20\%.

\section{Discussion }

\label{sec:Discussion}

\subsection{Strain distributions in XRD and HR-EBSD}

\label{subsec:DiscussStrains}

Both XRD profiles and probability distributions obtained from HR-EBSD
maps represent distributions of strains and rotations, as it was already
pointed out and compared by Wilkinson \emph{et al}. \citep{wilkinson14}.
Measurements by these two methods are complimentary. The XRD intensity
provides the probability distribution over several orders of magnitude
down from the maximum, allowing the dislocation density $\varrho$
and the dislocation strain screening distance $R$ to be determined
from a single profile. However, the XRD intensity is an average over
the sample volume and potentially includes a contribution from strained
regions far from the surface, in particular the film--substrate interface.
HR-EBSD, in turn, provides the maps of strains and rotations at the
surface. The statistics of the strain maps are limited by the number
of pixels in the map. The maps can be used to determine not only the
strain probability distributions, but also the spatial correlations
of strains and rotations, which allows the screening distance $R$
to be determined directly as the distance at which the correlations
become zero. A comparison of the dislocation correlations obtained
by the two methods is the main aim of the present paper.

We have compared in Fig.\,\ref{fig:XRD} the measured XRD curves,
their Monte Carlo simulations, and fits of both sets of curves by
the formula proposed in Ref.\,\citep{kaganer05GaN}. A comparison
of a curve simulated by the Monte Carlo method with the fit of this
curve provides a check on the internal consistency of simulations
and fits: dislocation densities obtained from fits of simulated curves
(open symbols in Fig.\,\ref{fig:XRD}(c)) coincide with the input
values of the simulations with fairly little scatter.

The dislocation densities that are obtained in the Monte Carlo simulations
and confirmed by the fits of the XRD curves are larger than the spot
densities obtained from Fig.\,\ref{fig:Samples} by factors of 2
and 3 for samples 1 and 2, respectively. It is noteworthy that our
previous studies \citep{kaganer05GaN,kopp13jac,kopp14jap} also yielded
dislocation densities obtained from the XRD profiles that were several
times larger than those determined by transmission electron microscopy.
Since XRD collects intensity scattered from the whole volume of the
GaN film, the additional broadening of XRD curves can be caused by
strain originating from the film--substrate interface. However, calculations
\citep{kopp14jap} show that misfit dislocations at the film--substrate
interface modify the reciprocal space maps but, for thick films, have
little effect on the double-crystal curves that are measured in the
present work.

\subsection{Screening distances}

\label{subsec:DiscussScreening}

The screening of dislocation strains by surrounding dislocations provides
a transition from the Gaussian central part of the XRD profile to
the $I(q)\propto q^{-3}$ tails. The screening distance $R$ is a
second parameter, in addition to the dislocation density $\varrho$,
that is included in the fit of the XRD profiles \citep{kaganer05GaN}.
The screening distances obtained by fitting the experimental curves
(full symbols in Fig.\,\ref{fig:XRD}(d)) agree with those obtained
from correlations in HR-EBSD maps (horizontal dashed lines in Fig.\,\ref{fig:XRD}(d))
within a factor of 2, with a rather large scatter between the values
obtained from different reflections.

Plots of the autocorrelation functions of the components of the strain
and rotation tensors measured by HR-EBSD provide a more robust tool
for determining the screening distance $R$. This distance is clearly
seen as a kink in the plot of the correlation function in dependence
of the logarithm of the separation, and is found to be the same for
all strain and rotation components. We find screening distances $R$
of 2\,\textmu m and 0.3\,\textmu m for the GaN films with threading
dislocation densities $\varrho$ of $5\times10^{8}$\,cm$^{-2}$
and $1.8\times10^{10}$\,cm$^{-2}$, respectively. In both cases,
the number of dislocations that provide screening of the strain of
a given dislocation $M=R\varrho^{1/2}$ is about 4. This means that
the dislocation strain field is very effectively screened by only
a few surrounding dislocations. To obtain these screening distances,
we measure the strain and rotation maps over areas with linear sizes
that are large compared to $R$ using steps that are small compared
to $R$. Different map sizes and steps are chosen depending on the
dislocation density. We note that very close numbers, 4 to 6 dislocations
providing screening of the dislocation strain fields, are obtained
for plastically deformed metals \citep{borbely00} by an XRD line
profile analysis method that is close to our approach for GaN films.

The screening distances are obtained directly by calculating the correlation
functions from the measured maps and do not require simulations. The
same analysis can be carried out for more complicated dislocation
distributions, when a simulation of the strain maps is not as straightforward
as in our case. We have performed simulations of the maps to better
understand them, but this is not necessary in every case where the
screening distance is of interest.

\subsection{Geometrically necessary dislocations}

\label{subsec:DiscussGND}

Maps of the strain and rotation components are commonly used in HR-EBSD
studies to determine the densities of geometrically necessary dislocations
(GND) \citep{sun00,eldasher03,pantleon08,adams09,wilkinson10,jiang13}.
The GND density corresponds to the minimum density of dislocations
that is required to provide a net effect that is of interest \citep{nye53,arsenlis99}.
It is therefore dependent on the scale of that effect and implies
an appropriate spatial averaging of the strains and rotations at smaller
scales. The papers cited above use the concept of GND to describe
boundaries between misoriented domains or plastic bending of a crystal.
The GND value is then a minimum density of dislocations that provides
a small angle boundary or the curvature of the crystal lattice. In
these cases, strain variations can be neglected compared to the lattice
rotations, and the GND can be obtained from the differences in rotations
between neighboring points in the measured maps.

\begin{figure*}
\centering \includegraphics[width=1\textwidth]{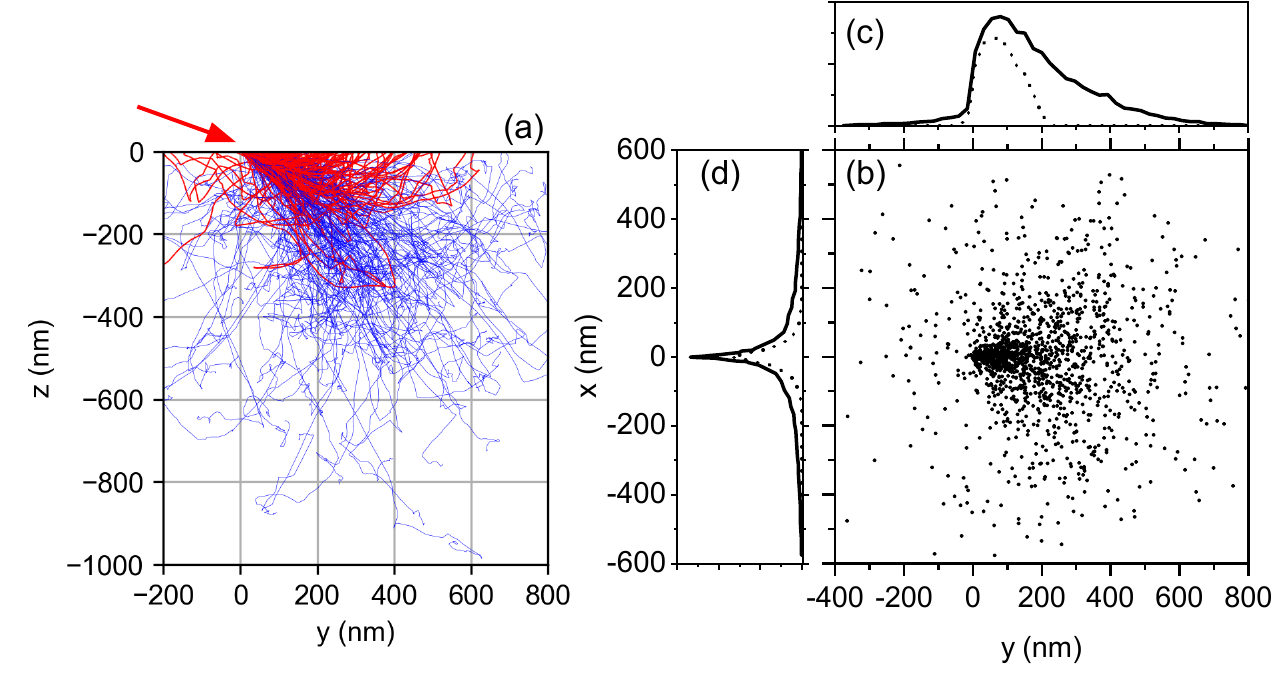}

\caption{(a) Side view of the electron trajectories in GaN simulated by CASINO,
(b) top view of the exit points of the trajectories of backscattered
electrons and (c,d) distributions of the exit points in projections
on the $x$ and $y$ axes. Dashed lines are the distributions of the
exit points of the electrons that lost less than 10\% of the initial
energy. The incident electron beam (shown by a red arrow) of energy
15 keV is inclined by 70$^{\circ}$ along the $y$ axis.}

\label{fig:Casino} 
\end{figure*}

Threading dislocations in GaN films are an example of a system where
the concept of GND requires a different consideration. The screening
distances $R$ that we have found provide a scale for the determination
of the GND. Counting dislocations on a smaller scale gives the density
of statistically stored dislocations, while the strain averaged over
distances exceeding $R$ can be treated as due to the GND. The GND
density for threading dislocations in GaN films is zero, since the
dislocations screen each other's strain fields and do not provide
a net long-range effect. More specifically, the strains and rotations
are zero after averaging over the areas exceeding the screening distance
$R$ in the respective sample. To obtain this zero average, the step
size of a measurement has to be small compared to $R$. In the studies
cited above, the GND density is obtained from the differences in rotations
between neighboring points without an average over the points that
would be measured in between. In our samples, strain and rotation
maps measured with a step larger than $R$ would only show an uncorrelated
random distribution with the same probability density as our strain
maps. The difference between neighboring measured points would not
contain any feasible information if the step of the measurements is
chosen larger than $R$. On the other end, a measurement with a step
small compared to $R$ would provide a local dislocation density that
is needed to generate the particular strain and rotation maps. This
latter treatment of GND is applied in Ref.\,\citep{vilalta17} to
a very similar case of threading dislocations in (In,Al)N films.

\subsection{Spatial resolution of HR-EBSD maps}

\label{subsec:DiscussResolution}

We find that the correlation functions deviate from the $\propto\ln r$
law at small distances and become more gradual. Also, the HR-EBSD
maps of sample 2 with high dislocation density measured with a step
between the measuring points of only 20\,nm, show a clear distinction
between the $x$ and $y$ directions. A comparison of the correlations
in the $x$ and $y$ directions and Monte Carlo simulations allow
us to estimate the spatial resolution of the HR-EBSD maps to be $20\times200$\,nm$^{2}$,
with the worse resolution along the direction of the tilt of the incident
electron beam. In metals, grain boundaries are used for a direct measurement
of the resolution, which depends on the material and the accelerating
voltage (we refer to the papers \citep{steinmetz10,tripathi19} which
also provide literature reviews). The resolutions in $x$ and $y$
directions are obtained by appropriately orienting the grain boundary
with respect to the direction of the electron beam. The resolution
that we obtain is comparable to that obtained in relatively light
metals but shows a larger anisotropy between the $x$ and $y$ directions.

For additional insight, we simulate the trajectories of the electrons
in GaN using the free software CASINO \citep{drouin07}. Figure \ref{fig:Casino}(a)
is a side view on the electron trajectories. Red lines are trajectories
of the backscattered electrons. A top view of their exit points is
shown in Fig.\,\ref{fig:Casino}(b). Their spatial distribution is
very asymmetric with respect to the $x$ and $y$ directions. The
projections of the outcrop distributions on the $x$ and $y$ axes
in Figs.\,\ref{fig:Casino}(c,d) are even broader than the resolution
that we obtain from the simulations of the strain and rotation maps.
Since low-loss electrons provide the major contribution to HR-EBSD
patterns \citep{deal08}, we plot by dotted lines in Figs.\,\ref{fig:Casino}(c,d)
the distributions of electrons that lost less than 10\% of their initial
energy of 15\,keV. The widths of these distributions are consistent
with the resolution that we found.

The HR-EBSD maps of the dislocation-free GaN sample 0 show only uncorrelated
random noise for all strain and rotation components. The widths of
these distributions are an order of magnitude larger compared to Si(001).
We also note that the XRD curves of this sample, albeit narrow compared
to the other samples, are significantly broader than expected for
its dislocation density \citep{kaganer15jpd}. The HR-EBSD measurements
on GaN sample 0 were performed before and after surface cleaning,
which had a negligible effect on the strain distributions. Hence,
neither surface roughness nor contamination do appear to be responsible
for the apparent strain in this sample. 

In the absence of dislocations, the remaining source of strain are
point defects, including impurities as well as native defects. The
density of native defects in GaN is orders of magnitude larger than
in Si. The density of Ga vacancies alone in GaN ranges from $10^{16}$--$10^{18}$\,cm$^{-3}$
\citep{saarinen97,saarinen98,oila03}, as compared to a vacancy density
below $1\times10^{14}$\,cm$^{-3}$ in Si \citep{voronkov99,bettin19}.
To examine the effect of point defects, we have performed simulations
of their strain maps for a spatially random distribution of dilatation
centers with the excess volume equal to the atomic volume. Even for
a density of $10^{18}$\,cm$^{-3}$, the strain produced by these
centers is too low to account for the experimentally measured strain
in Fig.\,\ref{fig:mapsEps22}(a). Moreover, the simulated strain
maps exhibit the same anisotropy as seen in Fig.\,\ref{fig:mapsEps22}(c)
caused by the inclination of the incident electron beam.

The distributions of the shear strains $\varepsilon_{13}$ and $\varepsilon_{23}$
coincide in all three GaN samples and are therefore independent of
the dislocation density. These strains have to be zero due to the
stress free boundary conditions. Of the three zero stress conditions
at the surface $\sigma_{i3}=0$ ($i=1,2,3)$, one condition, $\sigma_{33}=0$,
is used in the processing of the Kikuchi patterns. The other two conditions
reduce to $\varepsilon_{13}=0$ and $\varepsilon_{23}=0$ for the
(0001) surface of a hexagonal crystal. The distributions of these
components thus provide an estimate of the sensitivity limit to dislocation
strains. We note that the maps of these components do not show any
correlations in the crystals with dislocations, in contrast to other
strain and rotation components that develop correlations. Thus, in
the case where a dislocation free crystal is not available as a reference,
the distributions of these shear strain components can provide an
estimate of the accuracy of the other strain and rotation maps.

\section{Summary}

The screening of the dislocation strains by the surrounding dislocations
is revealed by both fits of the XRD curves in reciprocal space and
the real space correlations in the HR-EBSD strain maps. Our study
is performed using a well-defined array of threading dislocations
in GaN(0001) as an example. It allowed Monte Carlo simulations of
XRD curves and HR-EBSD maps within one and the same model of dislocation
distributions and its quantitative comparison with the experiment.
The simulations provide a basis for determining of the screening distances
from the HR-EBSD maps as the positions of the kinks in the strain
autocorrelations plotted against the logarithm of the distance. Determining
the screening distance in this way does not require any simulations
or fits and provides a robust tool also for more complicated dislocation
ensembles, where simulations would be more difficult.

In the XRD, dislocation densities and screening distances are obtained
by fitting the diffraction profiles. We find that XRD overestimates
the dislocation densities, presumably due to contributions from the
strained region near the film--substrate interface. The screening
distances obtained by XRD from different reflections possess larger
scattering and are consistent with those obtained by HR-EBSD within
a factor of two.

Screening distances of 2\,\textmu m and 0.3\,\textmu m are obtained
for the GaN films with threading dislocation densities of $5\times10^{8}$\,cm$^{-2}$
and $1.8\times10^{10}$\,cm$^{-2}$, respectively. This indicates
that the dislocation strain is screened by only four neighboring dislocations
in both samples. The strain maps from the high dislocation density
sample measured with a small step size show a large anisotropy in
the resolution of the HR-EBSD measurements. The resolution is estimated
to be $20\times200$\,nm$^{2}$, with the worse resolution in the
direction corresponding to the inclination of the incident electron
beam.

\section*{Acknowledgments}

The authors thank Bernd Jenichen for XRD measurements on sample 2
made for our former study \citep{kaganer15jpd}, Carsten Richter for
useful discussions, Jingxuan Kang for cleaning sample 0 and Moritz
Hansemann for AFM measurements of that sample, as well as Achim Trampert
for a critical reading of the manuscript. Arno Winkelmann and Graham
Meaden read the preprint of the present paper and analyzed our HR-EBSD
data with their software. We are grateful to them for their efforts
and comments.

\bibliographystyle{elsarticle-num}

\newpage

\section*{\large \center Supplementary material to the paper  \\ Dislocation correlations in GaN epitaxial films \\ revealed by EBSD and
XRD}

{\center by}

{\center Vladimir M. Kaganer, Domenik Spallek, Philipp John, Oliver Brandt, and Jonas Lähnemann}

\global\long\def\thefigure{SM\arabic{figure}}%
\setcounter{figure}{0}

\global\long\def\thesection{SM\, \arabic{section}}%
\setcounter{section}{0}

\section{Kikuchi patterns}

\begin{figure*}[h]
\centering \includegraphics[width=0.6\textwidth]{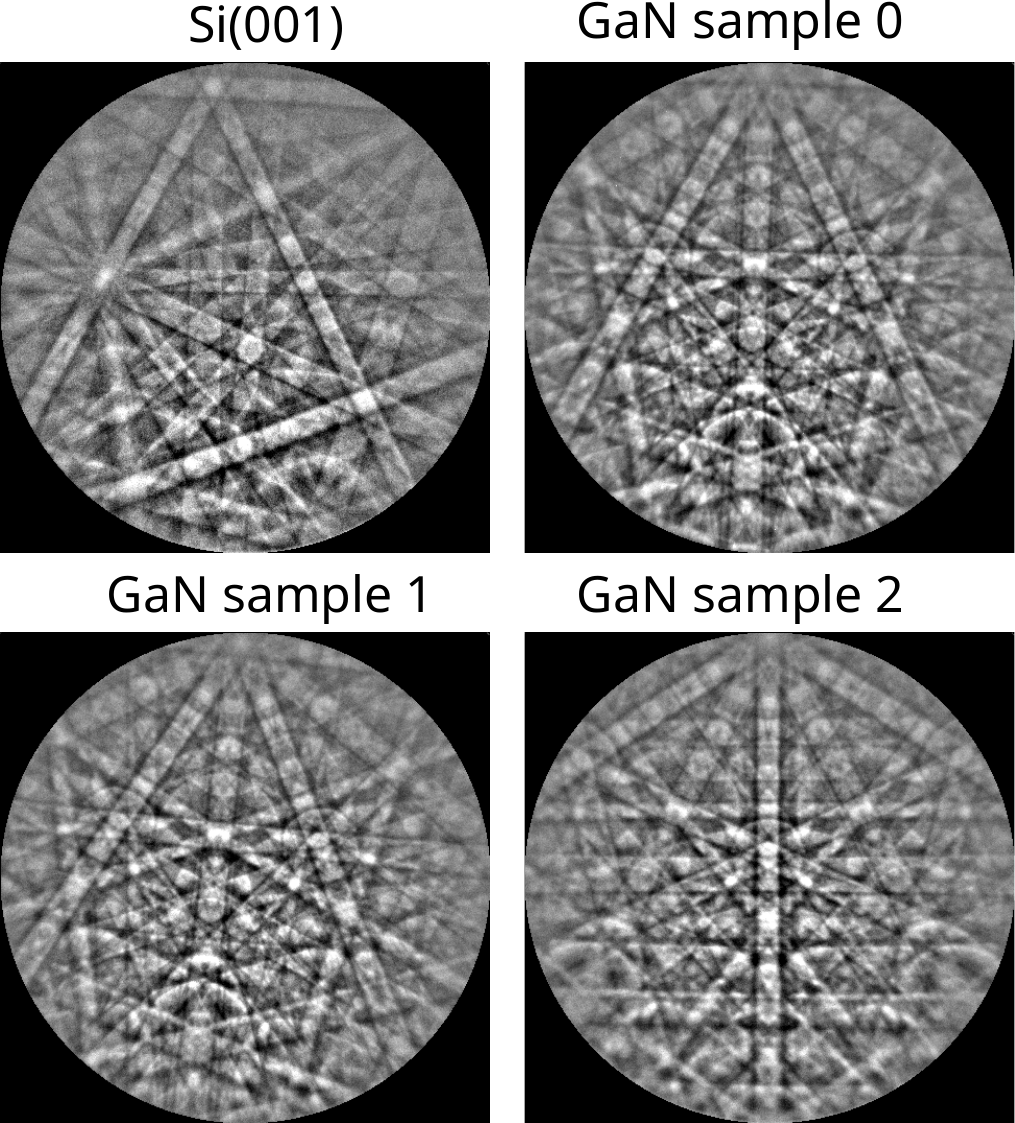}\caption{Kikuchi patterns of the GaN samples 0, 1, 2 and the Si wafer.}
\label{fig:Kikuchi} 
\end{figure*}

\newpage

\section{Strain and rotation maps}

\begin{figure*}[h]
\centering \includegraphics[width=1\textwidth]{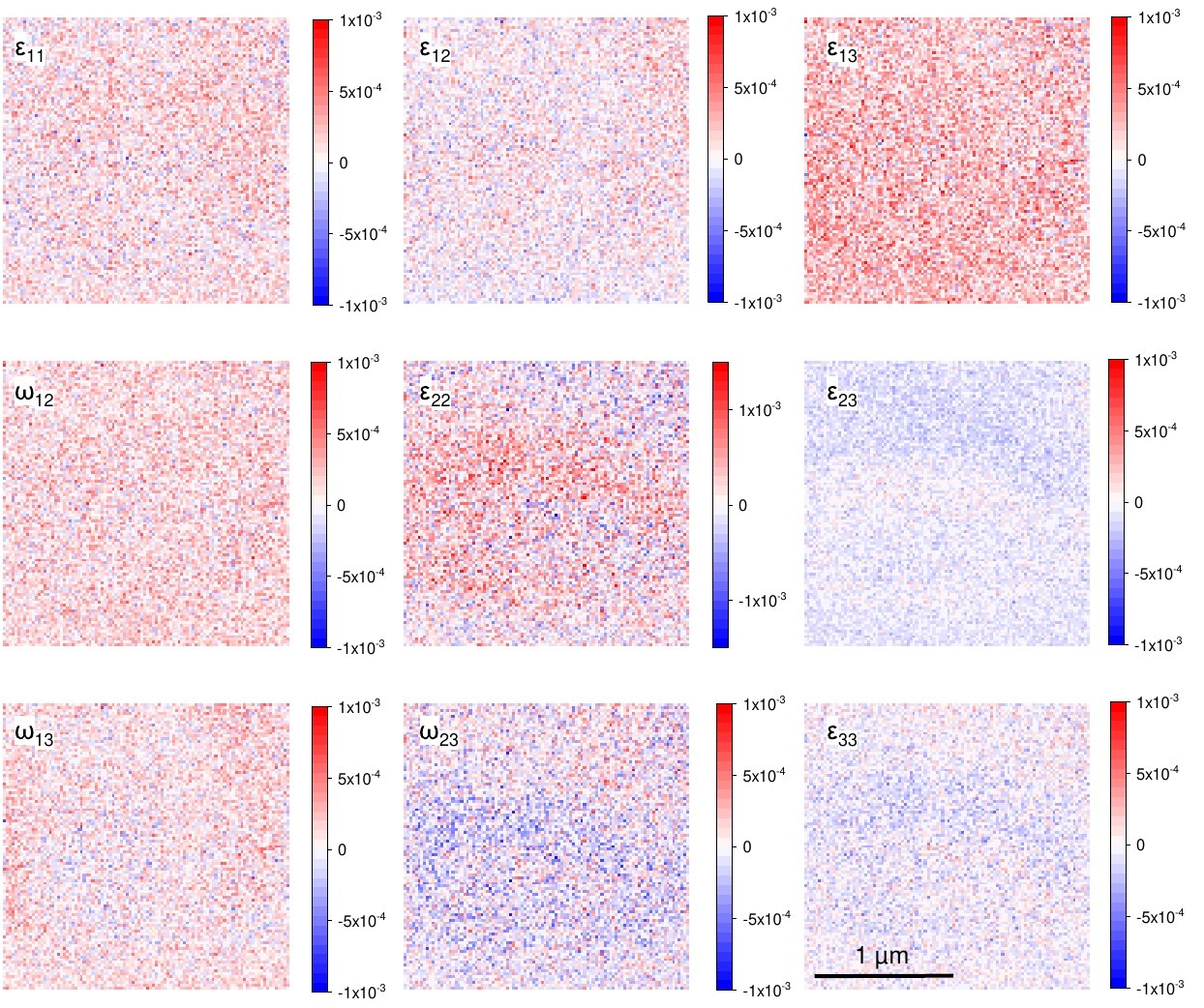}\caption{Components of the strain and rotation tensors of sample 0 (dislocation
free GaN sample).}
\label{fig:MapsSample0} 
\end{figure*}

\begin{figure*}
\centering \includegraphics[width=1\textwidth]{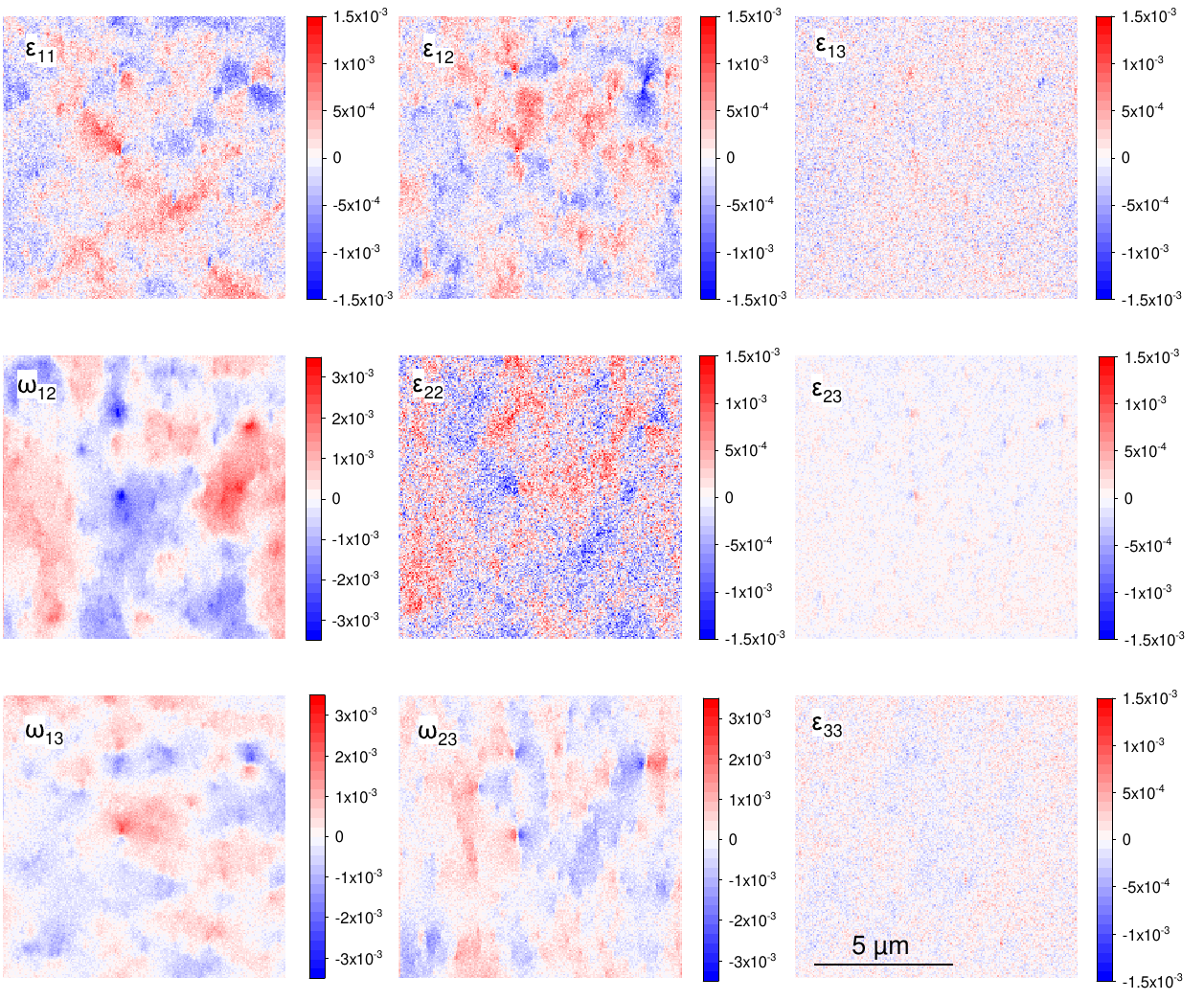}

\caption{Components of the strain and rotation tensors of sample 1.}

\label{fig:MapsSample1} 
\end{figure*}

\begin{figure*}
\centering \includegraphics[width=1\textwidth]{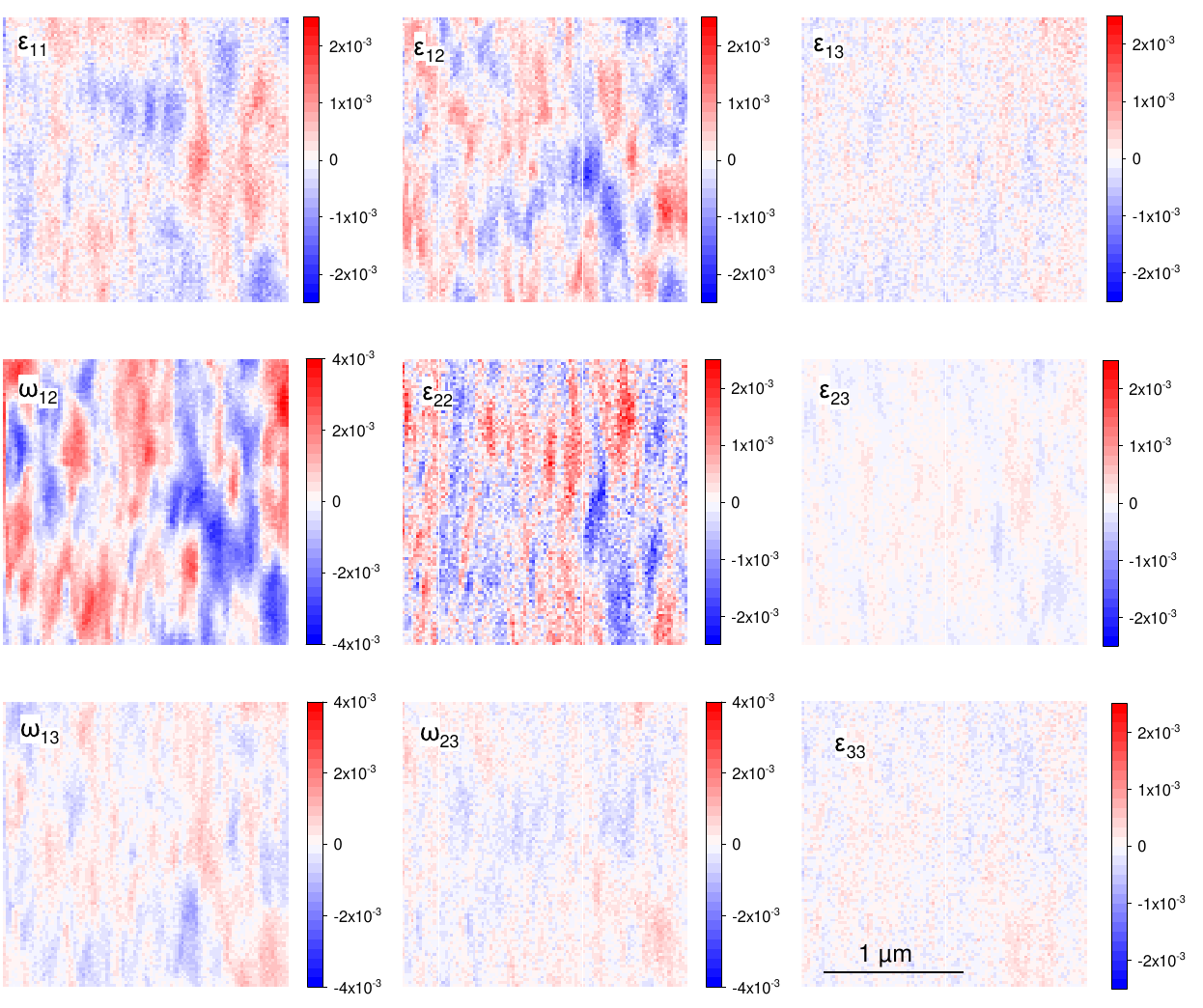}

\caption{Components of the strain and rotation tensors of sample 2.}

\label{fig:MapsSample2} 
\end{figure*}

\begin{figure*}
\centering \includegraphics[width=1\textwidth]{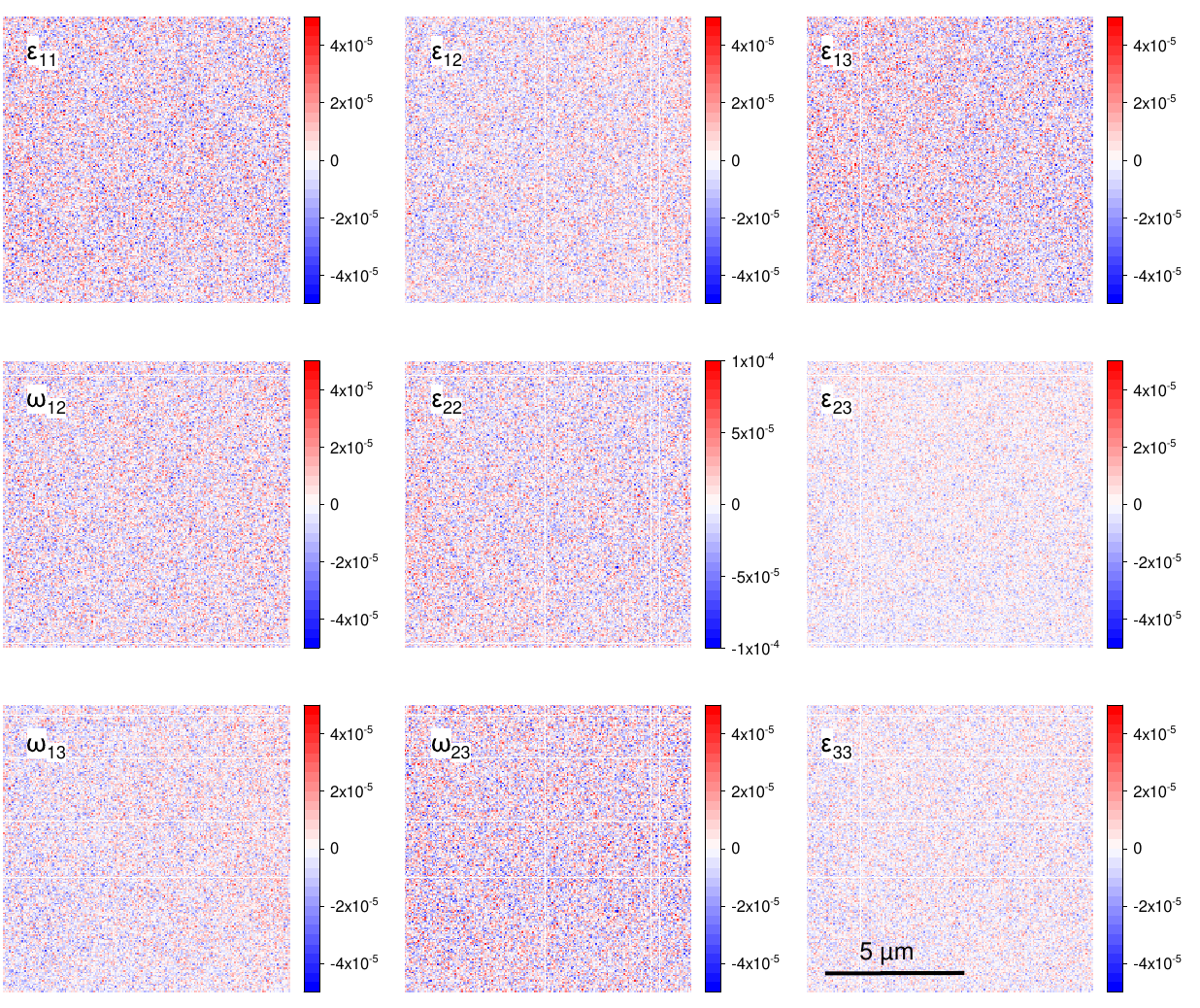}\caption{Components of the strain and rotation tensors of a silicon wafer.
Note different ranges of the values in comparison to Fig.\,\ref{fig:MapsSample0}.}
\label{fig:MapsSiliconWafer} 
\end{figure*}

\begin{figure*}
\centering \includegraphics[width=1\textwidth]{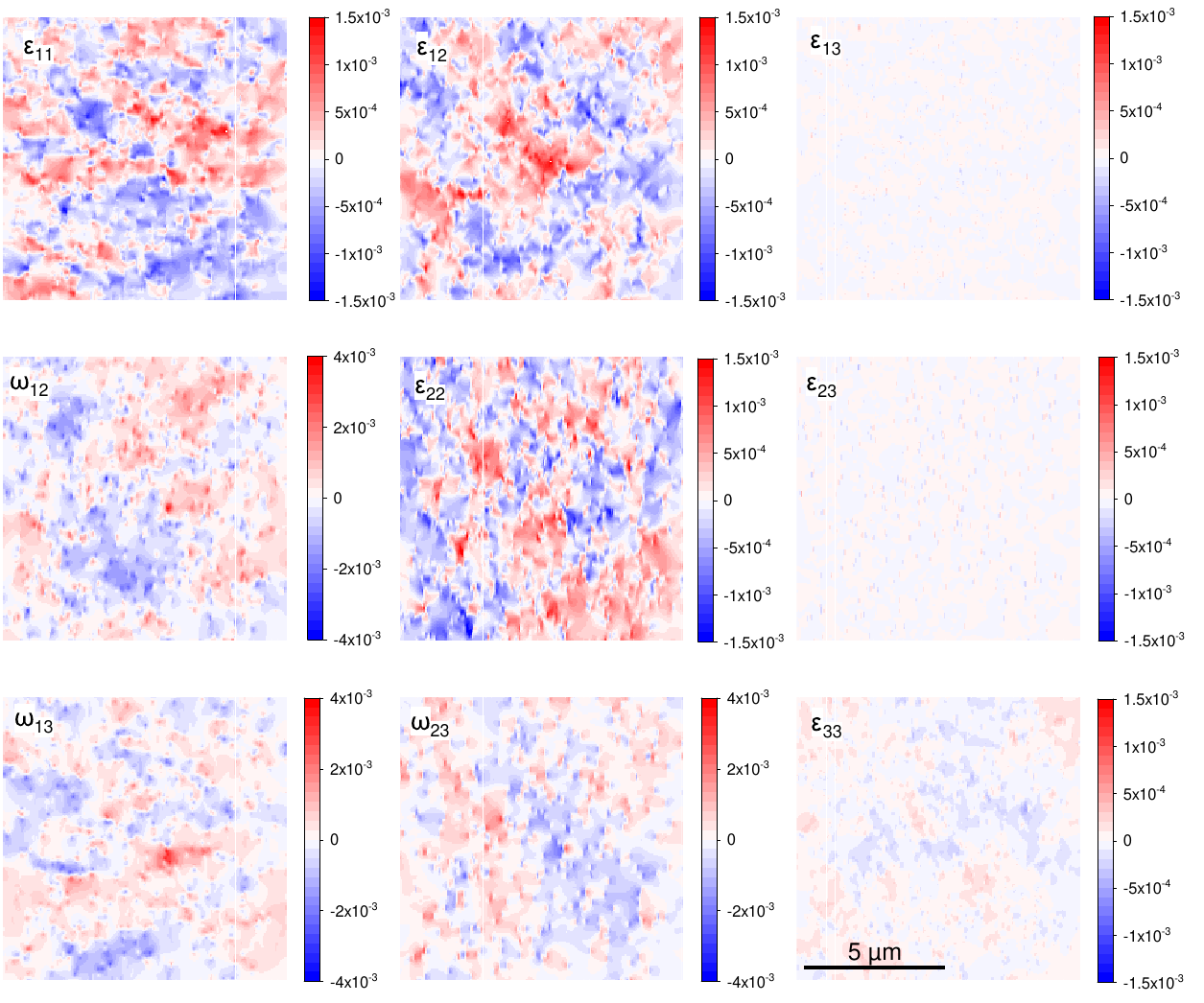}\caption{Monte Carlo simulation of the strain and rotation tensors of sample
1.}
\label{fig:MCsample1} 
\end{figure*}

\begin{figure*}
\centering \includegraphics[width=1\textwidth]{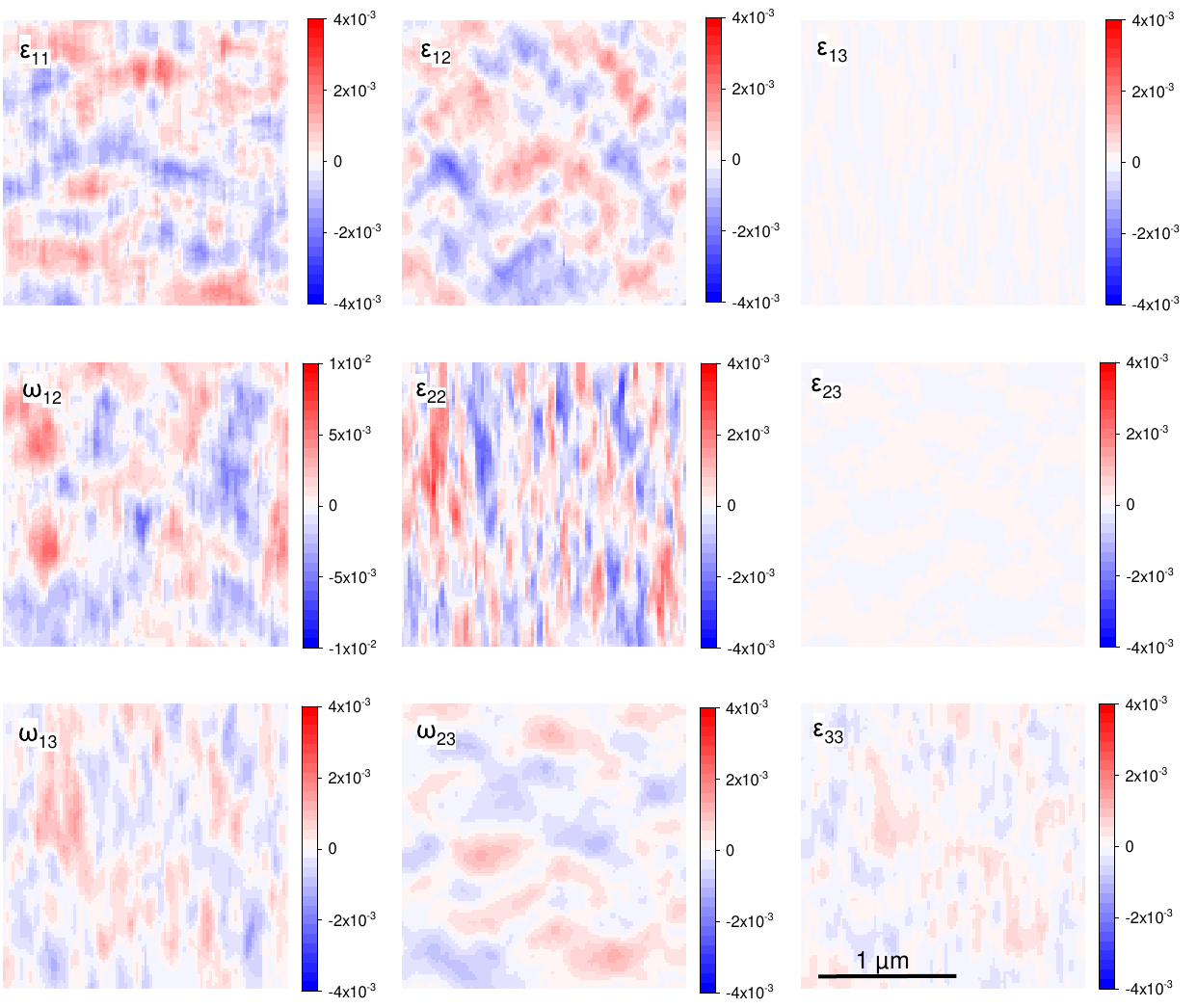}\caption{Monte Carlo simulation of the strain and rotation tensors of sample
2.}
\label{fig:MCsample2} 
\end{figure*}

\newpage

\section{Probability distributions and correlation functions of rotations}

\begin{figure*}[h]
\centering \includegraphics[width=1\textwidth]{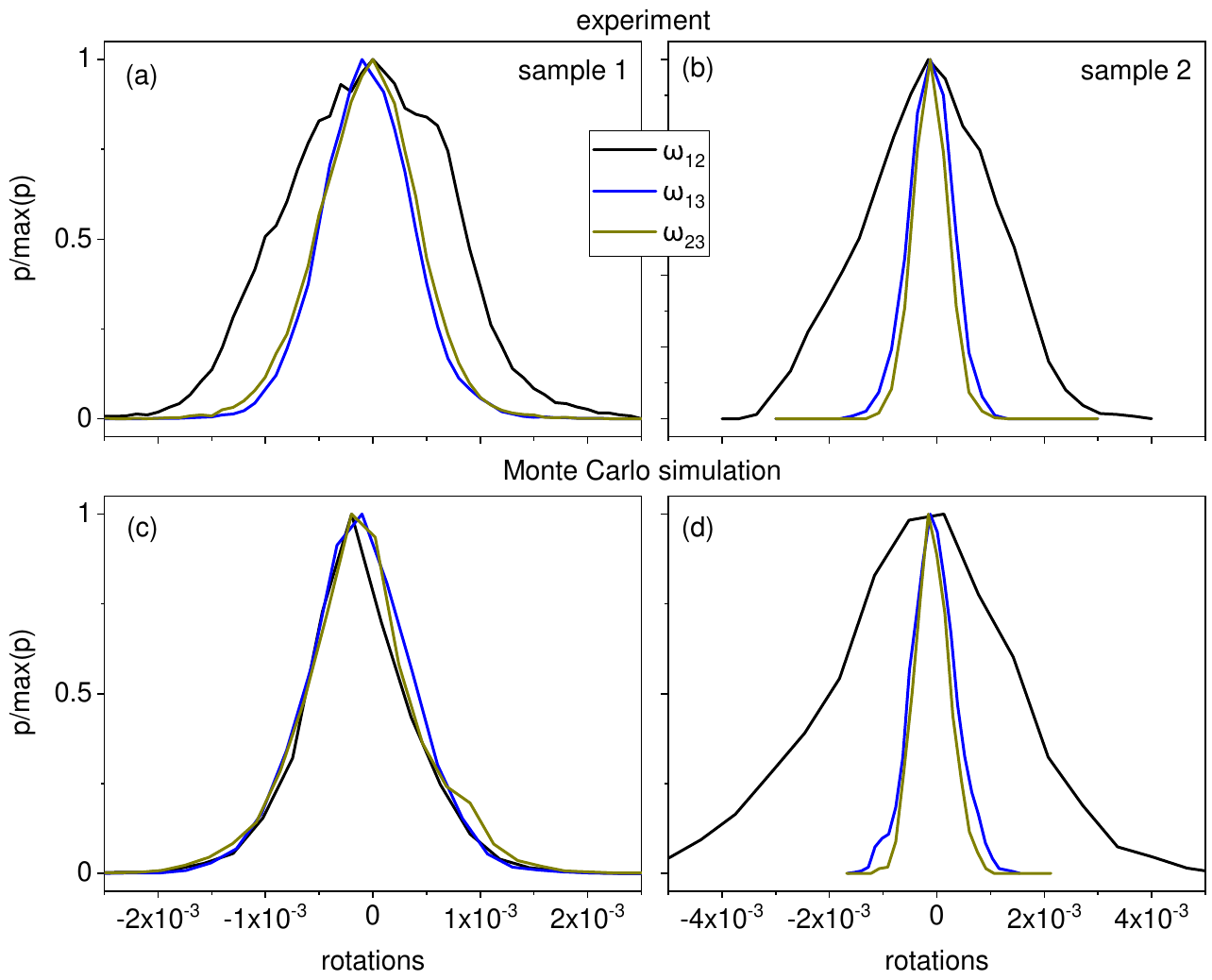}
\caption{Probability distributions of rotations (a,b)
in the measured maps and (c,d) in Monte Carlo simulations.}
\label{fig:HistogramsRotations} 
\end{figure*}

\begin{figure*}[h]
\centering \includegraphics[width=1\textwidth]{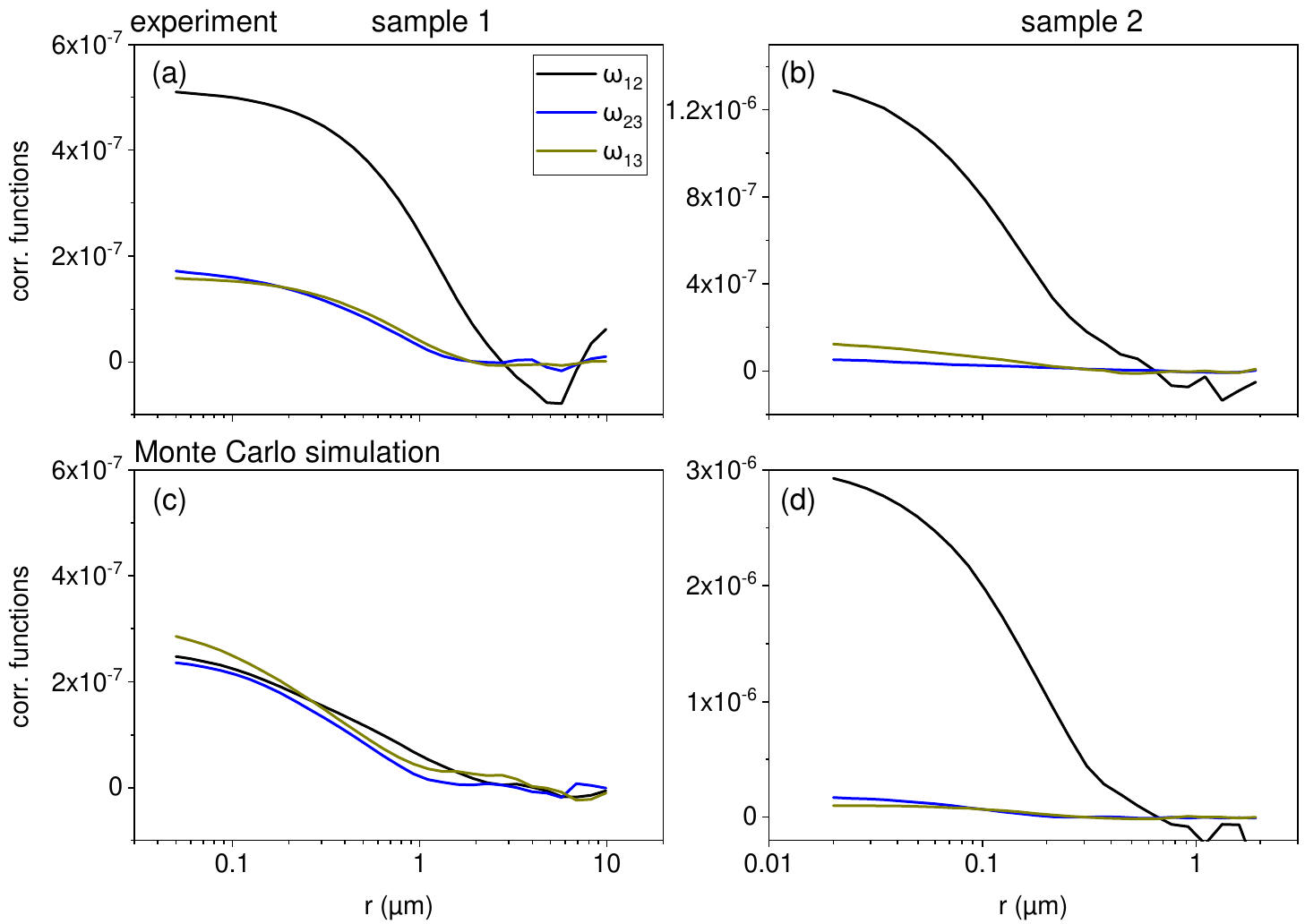}
\caption{Orientation-averaged autocorrelation functions of the rotations (a,b)
in the measured maps and (c,d) in Monte Carlo simulations.}
\label{fig:CorrRotations} 
\end{figure*}

\newpage

\section{Effect of drift}

\begin{figure*}[h]
\centering \includegraphics[width=1\textwidth]{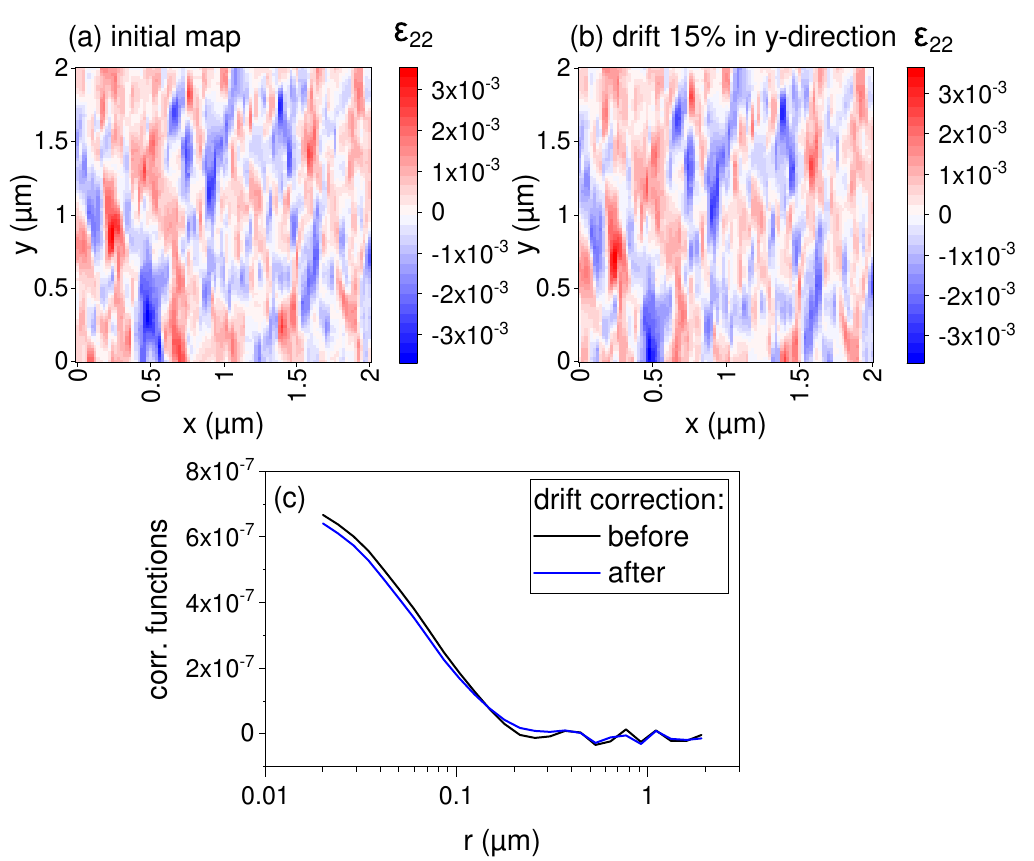}
\caption{(a) Monte Carlo simulated map of the strain $\varepsilon_{22}$  of sample 2 (identical to Fig.\ 3(f) in the main text and Fig.\ \ref{fig:MCsample2}), (b) the same map with
a 15\% drift in $y$-direction added, and (c) autocorrelation functions obtained from these maps. }
\label{fig:Drift} 
\end{figure*}

Figure \ref{fig:Drift} simulates the effect of drift on the correlation functions of strain. Figure \ref{fig:Drift}(a) is the Monte Carlo simulated map of the strain component  $\varepsilon_{22}$ of sample 2. The same map is presented in  Fig.\ 3(f) in the main text and in Fig.\ \ref{fig:MCsample2}. In Fig.\ \ref{fig:Drift}(b), a 15\% drift in $y$-direction is simulated, assuming that the effect of drift is accumulated linearly: each pixel of the map is shifted vertically proportional to the time between the measurement of that pixel and the first pixel, so that the last pixel is shifted by 15\%. Note that the drift effect is negligible at the top of the map and continuously increases to the bottom. The bottom part of the expanded map is cut out, to have the same area as in the initial map. Figure \ref{fig:Drift}(c) compares the autocorrelation functions obtained from these two maps. Since the correlations tend to zero at distances exceeding 0.2\thinspace\textmu m, which is 0.1 of the map size, only correlations of strains in closely spaced points are of interest. As a result, the effect of drift on the autocorrelation function in Fig.\ \ref{fig:Drift}(c) is negligibly small.

\newpage

\section{Repeated measurements and Monte Carlo simulations of EBSD maps}

\begin{figure*}[h]
\centering \includegraphics[width=1\textwidth]{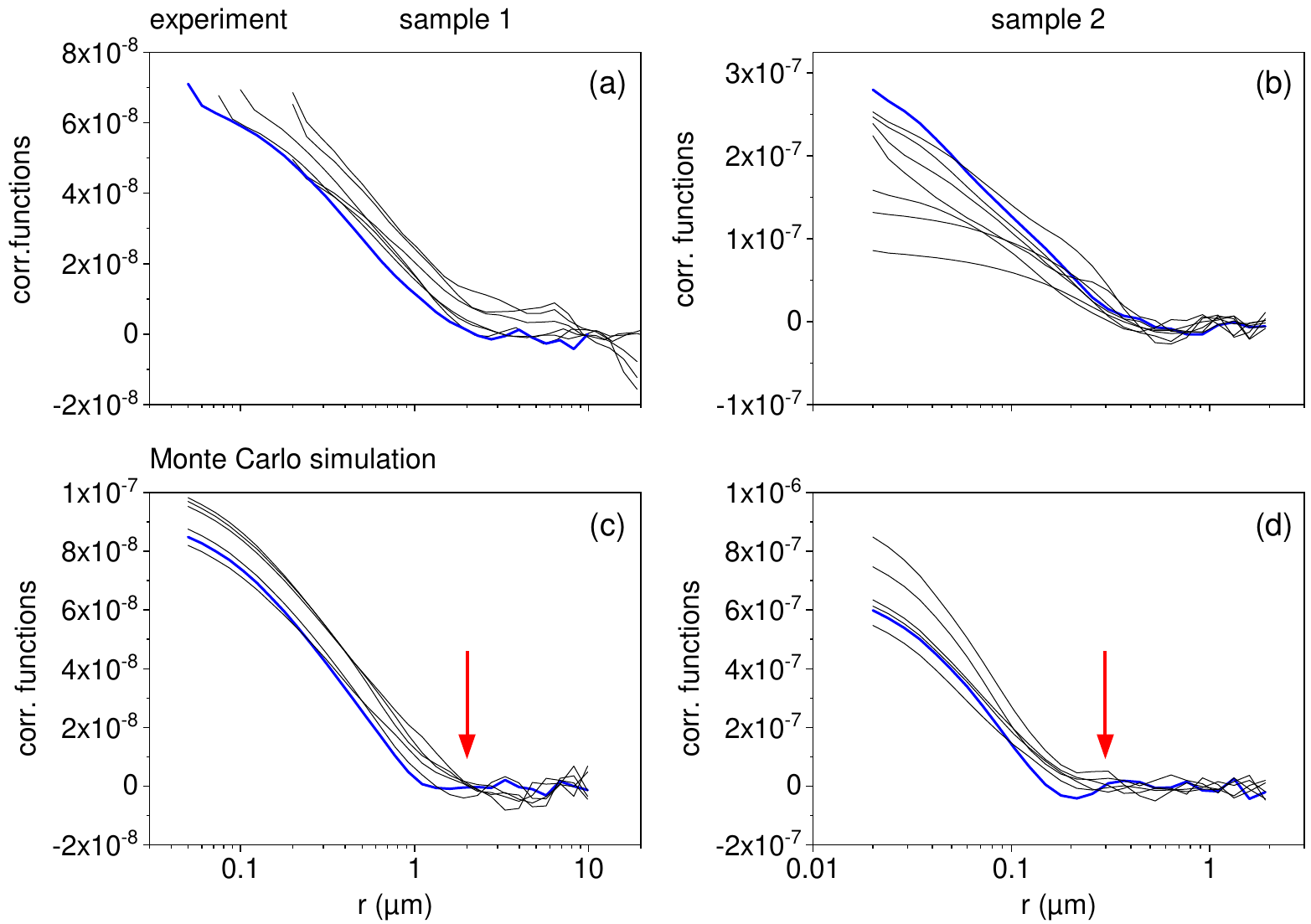}
\caption{Autocorrelation functions of the strain component $\varepsilon_{22}$ obtained (a,b) in the HR-EBSD study of samples 1 and 2 and (c,d)  by repeating Monte Carlo simulations of HR-EBSD maps  with all parameters kept the same as in Figs.\ 3--6. Blue lines are the calculations in Fig.\ 6. Black lines in the experimental plots are due to the measurements in different places of the respective samples, while black lines in the Monte Carlo modeling are obtained by changing the seed of the random number generator. The vertical red arrows point out to the screening distances taken on input of the Monte Carlo simulations (2 \textmu m and 0.3 \textmu m for samples 1 and 2 respectively). }
\label{fig:repetitions} 
\end{figure*}

\newpage

\section{Corrigendum to ``Dislocation correlations in GaN epitaxial films revealed by 
EBSD and XRD'' [Acta Materialia 297 (2025) 121357]}

The authors regret to report that an error was found in the processing
of the HR-EBSD maps of the Si wafer. The strains in this sample were
underestimated by a factor of approximately 60. The figure below reproduces
Fig.~4(a,b) from the paper, with the corrected curves added as thick
black lines. The scale of the strain map of this sample in Fig.~3(d), as well
as for all its strain and rotation maps in the Supplementary Material
(Fig.~SM5), requires the same correction. In consequence, HR-EBSD measurements
carried out on a Si wafer in the experimental setup used in the paper cannot be
used to determine the strain sensitivity of the experiment. As this
measurement on a Si wafer was only used for this purpose, correcting
the error does not affect any other results and conclusions in the
paper.

\begin{figure*}[h]
\centering \includegraphics[width=\textwidth]{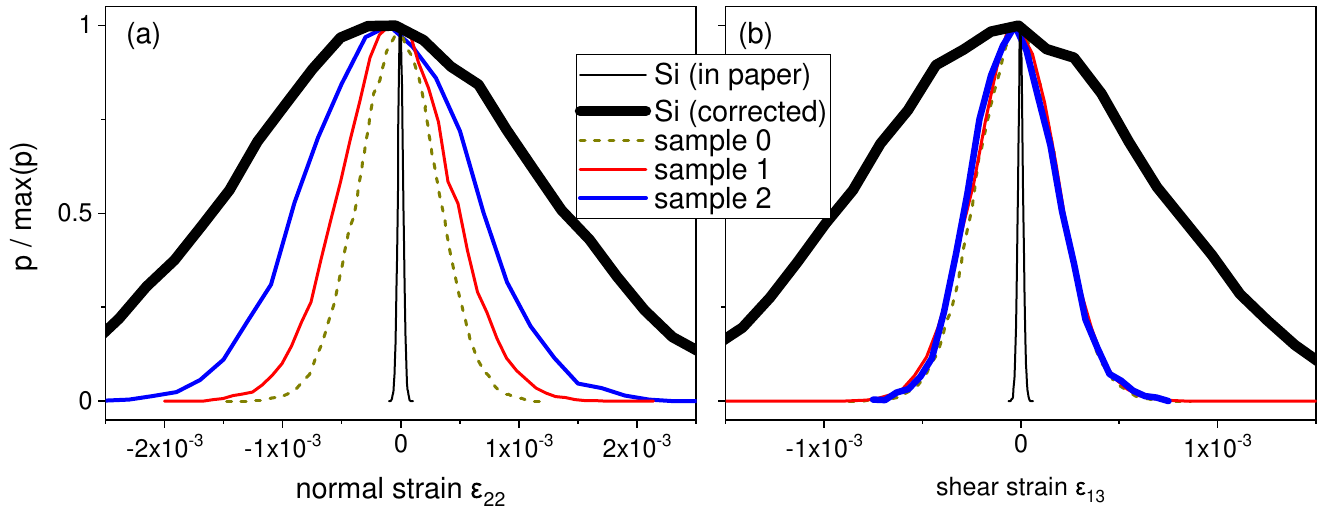}
\caption{Probabilities of (a) normal strain component $\varepsilon_{22}$ and
(b) shear strain component $\varepsilon_{13}$ in samples 1 and 2,
as well as in the reference dislocation free GaN sample (sample 0)
and a silicon wafer. The strains in the silicon wafer presented in
the paper are shown by thin black lines, while thick black lines are
the corrected curves.}
\end{figure*}

\end{document}